\documentclass[fleqn,usenatbib]{mnras}

\usepackage{amsmath}

\usepackage[T1]{fontenc}

\usepackage{xcolor}
\usepackage{soul}

\usepackage{comment}
\usepackage{cleveref}
\DeclareRobustCommand{\VAN}[3]{#2}
\let\VANthebibliography\thebibliography
\def\thebibliography{\DeclareRobustCommand{\VAN}[3]{##3}\VANthebibliography}

\usepackage{graphicx}	% Including figure files
\usepackage{amsmath}	% Advanced maths commands
\usepackage{nccmath}
\usepackage{mathtools}

\usepackage{booktabs,chemformula}
\usepackage{caption}% <-- added
\usepackage{array,booktabs,longtable,tabularx}
\usepackage{tabularray}
\usepackage{ltablex}% <-- added
\usepackage{siunitx}% <-- added
\usepackage{caption}% <-- added
\usepackage{fancyhdr}
\usepackage{tocloft}
\usepackage{verbatim}
\usepackage{multicol}
\usepackage{amsmath}
\usepackage{amssymb}
\usepackage{engtlc}
\usepackage{soulutf8}
\usepackage{multirow}
\usepackage{mathtools}
\usepackage{cancel}
\usepackage{subfig}
\usepackage{float}
\usepackage{empheq}
\usepackage[autostyle,italian=guillemets]{csquotes}
\usepackage[normalem]{ulem}

\newcommand{\secref}[1]{\S\ref{#1}}
\usepackage{newtxtext,newtxmath}

\title[]{Variety of disk wind-driven explosions in massive rotating stars}

\author[Crosato Menegazzi, L, et al.]{
Ludovica Crosato Menegazzi,$^{1}$
Sho Fujibayashi,$^{1}$
Koh Takahashi,$^{2,1}$
Ayako Ishii$^{3,1}$
\\
$^{1}$Max Planck Institute for Gravitational Physics (Albert Einstein Institute), Am M{\"u}hlenberg 1, Potsdam 14476, Germany\\
$^{2}$National Astronomical Observatory of Japan, National Institutes for Natural Science, 2-21-1 Osawa, Mitaka, Tokyo 181-8588, Japan\\
$^{3}$Faculty of Science, Yamagata University, 1-4-12 Kojirakawa-Machi, Yamagata, Yamagata 990-8560, Japan\\
}

\date{Accepted 2024 February 16. Received 2024 January 26; in original form 2023 November 13}

\pubyear{2024}

\begin{document}
\label{firstpage}
\pagerange{\pageref{firstpage}--\pageref{lastpage}}
\maketitle

\begin{abstract}
We perform a set of two-dimensional, non-relativistic, hydrodynamics simulations for supernova-like explosion associated with stellar core collapse of rotating massive stars to a system of a black hole and a disk connected by the transfer of matter and angular momentum. Our model of the central engine also includes the contribution of the disk wind. 
%In this work, we investigate the wind-driven explosion of rotating, large-mass progenitor stars with the zero-age main-sequence mass of $M_\mathrm{ZAMS}=20\,M_\odot$.
This study is carried out using the open-source hydrodynamic code \texttt{Athena++}, for which we implement a method to calculate self-gravity for axially symmetric density distributions.
We investigate the explosion properties and the $^{56}$Ni production of a star with the zero-age main-sequence mass of $M_\mathrm{ZAMS}=20\,M_\odot$ varying some features of the wind injection.
We find a large variety of explosion energy with $E_\mathrm{expl}$ ranging from $\sim 0.049\times10^{51}$~erg to $\sim 34\times10^{51}$~erg and ejecta mass $M_\mathrm{ej}$ from 0.58 to 6 $M_\odot$, which shows a bimodal distribution in high- and low-energy branches.
We demonstrate that the resulting outcome of a highly- or sub-energetic explosion for a certain stellar structure is mainly determined by the competition between the ram pressure of the injected matter and that of the infalling envelope. 
In the nucleosynthesis analysis the $^{56}$Ni mass produced in our models goes from $< 0.2~M_\odot$ in the sub-energetic explosions to $2.1~M_\odot$ in the highly-energetic ones. These results are consistent with the observational data of stripped-envelope and high-energy SNe such as broad-lined type Ic SNe.
%However, we find a tighter correlation between the explosion energy and the ejecta mass than that observationally measured. 

\end{abstract}


%%%%%%%%%%%%%%%%%%%%%%%%%%%%%%%%%%%%%%%%%%%%%%%%%%

%%%%%%%%%%%%%%%%% BODY OF PAPER %%%%%%%%%%%%%%%%%%

\section{Introduction}

Gamma-ray bursts (GRBs) are extragalactic cosmological sources of gamma-rays and among the most energetic events that we can observe in the Universe. They are short (GRBs typically last from few milliseconds up to a few minutes) and very intense flashes of gamma-rays of variable intensity, with fluxes up to $\sim100$ photons cm$^{-2}$s$^{-1}$ usually ranging from hundreds of keV up to $\sim$1 MeV. GRBs release gamma-rays reaching a total \textit{isotropic equivalent}  radiation energy (i.e. the radiated energy if the GRB was equally bright in all directions) of $\sim 10^{53}-10^{54}$ erg. GRBs properties, such as the total energy, spectra, and duration, can be useful source of information about their progenitor (\citealt{Meszaros_2006}, \citealt{Woosley_Bloom_2006}).

 Over the years evidence has showed that GRBs of the ``long-soft'' variety (lGRBs)  are likely to originate from the deaths of massive stars (\citealt{Woosley_1993}, \citealt{Woosley_2006}, \citealt{Woosley_Bloom_2006}, \citealt{Janiuk_2013}) and many gamma ray bursts have been now associated with bright supernovae (SNe) (\citealt{Woosley_2006}).

The Photometric  and spectroscopic observations suggest that GRBs and  their SNe progenitor have aspherical features. The signature of a conical geometry of the bursts manifests itself as a broad-band break in the power-law decay of the GRB afterglow, known as ``jet break''. This break can be explained by relativistic beaming of light emitted by a decelerated relativistic jet (\citealt{2001_Frail}, \citealt{2004_Piran}) and it is predicted to be achromatic. %So far, the only SN Ic-BL associated with lGRB is SN~1998bw (\citealt{Woosley_1999}).

Several scenarios have been proposed to explain the GRBs and associated SNe (\citet{Woosley_1993}, \citet{2004_Piran}, \citet{Woosley_2006}, \citet{Rodriguez_Cano}, \citet{10.1093/mnras/stac613}). 
One of the most promising  scenarios is the \textit{collapsar} scenario. The collapse of the core of a massive star ($\gtrsim8\,M_\odot$) at the end of its hydrostatic evolution is the starting point for a complex sequence of events with many possible outcomes. Specifically, progenitors with an even higher mass ($>16 \,M_\odot$), as shown by \citet{Woosley_2006}, are likely to undergo a failed supernova and form a black hole (BH) with an accreting disk. It has been shown that in failed supernova the disk wind generated by viscous dissipation inside the accretion disk may naturally be a source of the SN energy (\citet{Woosley_1993}, \citet{macfadyen1999collapsars}, \citet{Popham_Woosley_1999}) with an explosion energy $E_\mathrm{expl}>10^{52}$ erg and it has been found to be rich in $^{56}$Ni (as shown by \citet{Hayakawa_2018}).  
Also recent numerical studies based on this scenario have confirmed that a large amount of $^{56}$Ni ($\ge 0.1\, M_\odot$) can  be synthesized in the outflow from the disk (e.g., \citealt{Just2022} and \citealt{Fujibayashi2023arxiv}).
In this scenario, the interaction between the new-born BH and the still accreting stellar material is the engine for the relativistic jets.

Another promising scenario for the GRBs and associated Type Ic-BL SNe is the so-called \textit{proto-magnetar} scenario, in which highly magnetized and fast-rotating proto-neutron star generates the relativistic outflow. In this scenario, rotation leads to global asymmetries of the shock wave, which translates into the formation of highly collimated, mildly relativistic bipolar outflow (known as \textit{MHD-driven supernova}) as shown by  \citet{Obergaulinger_Aloy_2017}, \citet{Obergaulinger_Aloy_2020}, \citet{Obergaulinger_Aloy_2021_1}, \citet{Obergaulinger_Aloy_2021_2}.     
In their study \citet{Grimmet_HNe_2021} investigate the production of $^{56}$Ni by performing hydrodynamics simulations based on this scenario. In their most energetic model (in which they measure an energy deposition rate $>10^{52}$ erg s$^{-1}$), they find large masses of ejected $^{56}$Ni ($>0.05-0.45\,M_\odot$) which is in good agreement with the ranges inferred from the light curves of SNe Ic-BL ($0.12-0.8\,M_\odot$ with median at $0.28\,M_\odot$ as measured by \citealt{Taddia_2019}). Therefore, both GRB formation models, the proto-magnetar and the  collapsar scenario, seem to be equally plausible at the current moment.

The different properties of the ejecta such as mass, composition, velocity, and geometry  strongly depend on the explosion mechanism of the SN.  A key to investigate the ejecta properties, is to study its asphericity. Since the progenitor stars have to be rapidly rotating, when they collapse in both previously mentioned scenarios, the resulting ejecta may naturally have aspherical features, as seen in observations of SN~1998bw. When the injection of the energy in the stellar envelope is  aspherical, the matter can keep infalling also after a successful explosion. As the energy source of the ejecta may be the infalling mass to the central engine, the feedback of the injected outflow on the infall stellar envelope is an important effect in the scenario. In the case of relativistic bipolar jets, the feedback effect have been extensively studied (\citealt{10.1093/mnras/stt2199}, \citealt{LIU20195}). However, such an effect by sub-relativistic outflow has not yet been studied in a systematic manner.

The motivation for this work is, therefore, to explore the properties of the ejecta based on the collapsar scenario, with a focus on the late-phase mass ejection after BH formation. We perform a set of two-dimensional hydrodynamics simulations of axisymmetric models of the ejecta generated by the collapse of rotating massive star. Based on the collapsar scenario, we assume that the explosion is powered through a BH-accretion disk system. We vary several parameters controlling the properties of the mass and energy injection to investigate their impact on the final ejecta.

The paper is structured as follows.  
We begin by explaining the hydrodynamic code we utilize in this work (the hydrodynamic equations it solves, the model for the central engine, the equation of state), 
the characteristics of the progenitor star we employ (taken from \citealt{Aguilera-Dena_2020}) and the setup for our simulations (inner boundary conditions and the free parameters of our models), in Section~\secref{sec:Method}. In Section~\secref{sec:Results} we present the results of the simulations focusing on the hydrodynamics of the explosion, the ejecta property and the $^{56}$Ni production with a systematic variation of the initial parameters. Here we also compare our results with  observational data and a general relativistic neutrino-radiation viscous-hydrodynamics simulation performed using the same progenitor from the literature. We discuss the implications of our results also considering the observational counterpart. We summarize this work in Section~\secref{sec:Conclusion}. The Appendixes provide the description of the multipole expansion of the gravitational potential we implement in our code and an insight to the model of the disk wind we used.

\section{Method}\label{sec:Method} 
We study the explosion of rotating massive stars ($M\sim 20\,M_\odot$) by performing a set of 2D non-relativistic simulations using the open-source multi-dimensional hydrodynamics code \texttt{Athena++} (\citealt{Stone_2020}). The nucleosynthesis calculation is performed at posteriori using the reaction network \texttt{torch} \citep{Timmes2000jul} on tracer particles.

\subsection{The scenario}
For this study, we consider the case of a failed core collapse supernova CCSN, in which the neutrino-driven explosion in the proto-neutron star (PNS) phase does not occur, leading the PNS to collapse into a BH. In this collapsar scenario we model the explosion of a compact progenitor star after the formation of a BH (but see \citet{Burrows_2020} for a different scenario). As a progenitor we employ the model provided by \citet{Aguilera-Dena_2020} of a rapidly rotating, rotationally mixed star with a $20M_\odot$ ZAMS mass. 
We then build a semi-analytical model for the central engine by taking into account the BH and disk evolution, which in this scenario are governed by the transfer of matter and angular momentum. This method is based on the prescriptions provided by \citet{Kumar_Narayan_method} on which we add the contribution of the disk following \citet{Hayakawa_2018}. 

We chose this progenitor because this kind of stars, with masses ranging from 4 to 45 $M_\odot$ were proposed to be progenitors of both superluminous SNe and long gamma-ray bursts (\citealt{Japelj_2016}, \citealt{Margalit_2018}, \citealt{Aguilera-Dena_2020}).  
We specifically employ the model with $M_\mathrm{ZAMS}=20\, M_\odot$ because it is supposed to fail the explosion (\citealt{Ertl_2016} and \citealt{Muller_2016}) as it has a very compact core with a core compactness of $\xi_{2.5}>0.6$.
%as shown by \citet{Aguilera-Dena_2020}.
%They computed this value
Here, the core compactness is calculated by
following \citet{OConnor2011}:

\begin{align}
    \xi_{M/M_\odot} = \frac{M/M_\odot}{R(M)/1000\,\mathrm{km}},
\end{align}

where $R(M)$ is the radius at which its enclosed mass is $M$, and it is measured at a mass coordinate of 2.5$M_\odot$ at the core collapse.
This quantity measures the gravitational binding energy near the core of pre-SN stars and is considered as an indicator of whether the collapse of a non-rotating stellar core leads to a successful explosion, or ends up with the formation of a BH instead.
\citet{Sukhbold_2014} found that, if $\xi_{2.5}>0.45$ at the core collapse, the core collapse is likely to fail the explosion and form a BH.
Therefore this $M_\mathrm{ZAMS}=20\, M_\odot$ progenitor well suits our central engine model in this sense.

\subsection{Hydrodynamic equations}
We perform two-dimensional non-relativistic hydrodynamic simulations using an open-source code \texttt{Athena++}.
In addition to the original functions, we newly implement the gravitational potential $\Phi$ by solving the Poisson's equation under the cylindrical symmetry.
The set of equations solved in this work is as follows:

%\begin{ceqn}
\begin{align}
   %\partial_t\rho + \nabla\cdot(\rho \mathbf{v}) &= 0 \\
   %&\partial_t\rho+\frac{1}{r^2}(r^2 \rho v_r)+\frac{1}{r\sin\theta}\partial_\theta (\sin\theta \rho v_\theta)+\frac{1}{r\sin\theta}\partial_\phi(\rho v_\phi) =0 \label{conitnuity_equation}\\
   &\partial_t\rho+\frac{1}{r^2}(r^2 \rho v_r)+\frac{1}{r\sin\theta}\partial_\theta (\sin\theta \rho v_\theta) =0 \label{conitnuity_equation}\\
    %\partial_t(\rho \mathbf{v})+\mathbf{\nabla}\cdot(\rho \mathbf{v}\mathbf{v}+P \mathbf{I}) &= -\rho\nabla\phi \label{momentum_cons}\\
    %&\partial_t(\rho v_r)+\frac{1}{r^2}\partial_r(r^2\rho v_r^2)+\frac{1}{r\sin\theta}\partial_\theta(\sin\theta v_rv_\theta)+ \notag \\
   % &\;\;+\frac{1}{r\sin\theta}\partial_\phi(\rho v_r v_\phi)-\rho\frac{v_r^2+v_\phi^2}{r}+\partial_rP=-\rho\partial_r \Phi \label{momentum_cons1},\\
    &\partial_t(\rho v_r)+\frac{1}{r^2}\partial_r(r^2\rho v_r^2)+\frac{1}{r\sin\theta}\partial_\theta(\sin\theta v_rv_\theta) \notag \\
    &\qquad\quad\, -\rho\frac{v_r^2+v_\phi^2}{r}+\partial_rP=-\rho\partial_r \Phi \label{momentum_cons1},\\ 
    %&\partial_t(\rho v_\theta)+\frac{1}{r^2}\partial_r(r^2\rho v_r v_\theta)+\frac{1}{r\sin\theta}\partial_\theta(\sin\theta\rho v_\theta^2)+\frac{\rho v_r v_\theta}{r}+ \notag \\ 
    %&\;\;+\frac{1}{r\sin\theta}\partial_\phi(\rho v_\theta v_\phi)-\frac{\cos\theta}{\sin\theta}\frac{\rho v_\phi^2}{r}+\frac{1}{r}\partial_\theta P=-\frac{\rho}{r}\partial_\theta\Phi \label{momentum_cons2},\\
    &\partial_t(\rho v_\theta)+\frac{1}{r^2}\partial_r(r^2\rho v_r v_\theta)+\frac{1}{r\sin\theta}\partial_\theta(\sin\theta\rho v_\theta^2) \notag \\ 
    &\qquad\quad\;\,+\frac{\rho v_r v_\theta}{r}-\frac{\cos\theta}{\sin\theta}\frac{\rho v_\phi^2}{r}+\frac{1}{r}\partial_\theta P=-\frac{\rho}{r}\partial_\theta\Phi \label{momentum_cons2},\\
   %& \partial_t(\rho v_\phi)+ \frac{1}{r^2}\partial_r(r^2\rho v_r v_\phi)+ \frac{1}{r\sin\theta}\partial_\theta(\sin\theta\rho v_\theta v_\phi)+\frac{\rho v_r v_\phi}{r} + \notag \\ 
   %&\;\;+\frac{\cos\theta}{\sin\theta}\frac{\rho v_r v_\phi}{r} +\frac{1}{r\sin\theta}\partial_\phi(\rho v_\phi^2 + P)= -\frac{\rho}{r\sin\theta}\partial_\phi \Phi\label{momentum_cons3},\\
   & \partial_t(\rho v_\phi)+ \frac{1}{r^2}\partial_r(r^2\rho v_r v_\phi)+ \frac{1}{r\sin\theta}\partial_\theta(\sin\theta\rho v_\theta v_\phi) \notag \\ 
   &\qquad\quad\;\,+\frac{\rho v_r v_\phi}{r} +\frac{\cos\theta}{\sin\theta}\frac{\rho v_r v_\phi}{r} = 0 \label{momentum_cons3},\\
   %\partial_t e_\mathrm{t} +\nabla\cdot[(e_\mathrm{t} +P)\mathbf{v}] &=-\rho \mathbf{v}\cdot\nabla\phi \label{energy_cons}
   %&\partial_te_\mathrm{t}+\frac{1}{r^2}\partial_r[(e_\mathrm{t}+P)v_r]+\frac{1}{r\sin\theta}\partial_\theta(\sin\theta(e_\mathrm{t}+P)v_\theta)+\notag\\
   %&+\frac{1}{r\sin\theta}\partial_\phi[(e_\mathrm{t}+P)v_\phi]=-\rho\Big(v_r\partial_r\Phi + v_\theta \frac{\partial_\theta \Phi}{r} +v_\phi\frac{\partial_\phi \Phi}{r\sin\theta} \Big)\label{energy_cons},
   &\partial_te_\mathrm{t}+\frac{1}{r^2}\partial_r[(e_\mathrm{t}+P)v_r]+\frac{1}{r\sin\theta}\partial_\theta(\sin\theta(e_\mathrm{t}+P)v_\theta)\notag\\
   &\qquad\qquad\qquad\qquad\qquad\qquad=-\rho\Big(v_r\partial_r\Phi + v_\theta \frac{\partial_\theta \Phi}{r} \Big)\label{energy_cons},
\end{align}
%\end{ceqn}
where $\rho$, $v_i$ (with $i=r$,  $\theta$ and $\phi$), $P$ and $e_\mathrm{t}$ are the density, the  velocity components, the pressure and the total energy density of the fluid respectively. $e_\mathrm{t}=e_\mathrm{kin}+e_\mathrm{int}$ is the sum of the kinetic energy density $e_\mathrm{kin}=(1/2)\rho v^2$ and the internal energy density $e_\mathrm{int}$. $\Phi$ is the gravitational potential which satisfies the Poisson's equation:

\begin{align}\label{Poisson_eq}
        \Delta \Phi=4\pi G\rho.
\end{align}

where $G$ is the gravitational constant, $\rho$ is the density, and $\Delta$ is the Laplacian. 
This system of equations shows the continuity equation~\eqref{conitnuity_equation}, 
the Euler's equation for the radial, latitudinal, and longitudinal components of the momentum (respectively equation~\eqref{momentum_cons1}, equation~\eqref{momentum_cons2}, and equation~\eqref{momentum_cons3}), and the energy equation~\eqref{energy_cons}.
We compute these equations using a finite volume method on a spherical grid.

\subsection{The gravity solver} \label{gravity_solver}
To evaluate the self-gravity in the spherical-polar coordinates, we implemented a gravitational potential solver in our code.
In the solver we first use the method of Green's function to obtain the integrated form of the gravitational potential $\Phi(\mathbf{r})$ as:

\begin{align}
        \Phi(\mathbf{r}) = -4 \pi G \int{\frac{\rho(\mathbf{r'})}{4\pi|\mathbf{r-r'}|}d\mathbf{r'}},
\end{align}

where $\mathbf{r}$ is a position vector. The potential $\Phi(\mathbf{r})$ also satisfies the Poisson's equation (equation~\eqref{Poisson_eq}).
To perform the integration, we use a multipole expansion described in \citet{Hachisu_gravitysolver}. We provide more details on the implementation of this solver in Appendix~\ref{Appendix:2D_gravity_solver}.

\subsection{The computational setup}

In this work we use an axisymmetric grid with spherical-polar coordinate. Our domain extends from 0 to $\pi$  for the polar dimension, and from $10^8$ cm ($r_\mathrm{in}$) to $3.3\times 10^{10}$ cm ($r_\mathrm{out}$ ) for the radial dimension. The inner radius determines the inner boundary inside which the enclosed mass is $1.28\;M_\odot$ and it roughly corresponds to the dimension of the iron core at core-collapse which we cut out from the computational domain. 
The outer radius extends over the stellar surface ($r_\mathrm{star} = 2.7\times 10^{10}\; \rm{cm}$).  The initial mass in the computational domain is $M_\mathrm{domain}=14.2 \,M_\odot$.

The computational domain is discretized by $128$ grid points uniformly in the $\theta$-direction and $220$ grid points with geometric spacing in the $r$-direction, in which the mesh size increases with a constant factor $\Delta r_i = \alpha \Delta r_{i-1}$. We chose the ratio as $\alpha\approx 1.03$ to ensure that all the meshes are approximately squared, i.e., $\Delta r_i \approx r_i\Delta \theta$. This grid resolution was chosen after a convergence study described in Appendix~\ref{Appendix:resolution_study}.

\subsection{The central engine model}\label{central_engine_model}
In our simulations, the computational domain does not contain the central engine, which is considered as being embedded in the central part of the star (at $r < r_\mathrm{in}$).
We evolve the masses of the disk and the BH $M_\mathrm{disk}$ and $M_\mathrm{BH}$ and their angular momenta $J_\mathrm{disk}$ and $J_\mathrm{BH}$ as follows:
\begin{align}
    \frac{dM_\mathrm{disk}}{dt} &= \dot{M}_\mathrm{fall,disk}-\dot{M}_\mathrm{acc}-\dot{M}_\mathrm{wind} \label{M_dot_disk},\\
    \frac{dM_\mathrm{BH}}{dt} &= \dot{M}_\mathrm{fall,BH}+\dot{M}_\mathrm{acc},\\
    \frac{dJ_\mathrm{disk}}{dt} &=  \dot{J}_\mathrm{fall,disk}-\dot{J}_\mathrm{acc}-\dot{J}_\mathrm{wind}, \label{J_dot_disk}\\
    \frac{dJ_\mathrm{BH}}{dt} &= \dot{J}_\mathrm{fall,BH}+\dot{J}_\mathrm{acc}, \label{J_dot_BH}
\end{align}
where $\dot{M}_\mathrm{fall,BH}$ and $\dot{M}_\mathrm{fall,disk}$ are the rates of the mass accretion that directly infalls onto the BH and onto the disk, respectively. $\dot{J}_\mathrm{fall,BH}$ and $\dot{J}_\mathrm{fall,disk}$ are the momentum accretions rates respectively associated to the BH and the disk.
We evaluate the fraction of the infalling matter that directly falls into the BH by considering the competition between the specific angular momentum of the infalling matter and that of the innermost stable circular orbit (ISCO) $j_\mathrm{ISCO}$. More precisely, if the specific angular momentum is smaller than $j_\mathrm{ISCO}$, the infalling mass accretes onto the BH, if instead it is larger than $j_\mathrm{ISCO}$, it becomes a part of the disk.
To determine $j_\mathrm{ISCO}$, we follow the prescription of \citet{Bardeen_ISCO}. We first evaluate the BH spin parameter $a$: 
\begin{align}
    &a(t) = \frac{cJ_\mathrm{BH}(t)}{GM_\mathrm{BH}^2(t)},
\end{align}
we then compute the ISCO radius, $r_\mathrm{ISCO}$, in terms of $a$ following:
\begin{equation}
    r_\mathrm{ISCO}(t)=\frac{GM_\mathrm{BH}(t)}{c^2}\left(3+z_2-\sqrt{(3-z_1)(3+z_1+2z_2)} \right),
\end{equation}
where $z_1$ and $z_2$ are given by:
\begin{align}
    z_1(t) & = 1+(1-a^2(t))^{1/2} \left((1+a(t))^{1/3}+(1-a(t))^{1/3} \right),\\
    z_2(t) &= (3a^2 +z_1^2)^{1/2}.
\end{align}
We then define the specific angular momentum at the ISCO at the first order  as follow:
\begin{equation}\label{j_isco}
    j_\mathrm{ISCO} \approx \sqrt{GM_\mathrm{BH}(t)r_\mathrm{ISCO}(t)}.
\end{equation}
The different accretion rates are then estimated at the inner boundary $r=r_\mathrm{in}$ as:
\begin{align}
    &\dot{M}_\mathrm{fall,BH} = 2\pi r_\mathrm{in}^2\int_{-1}^1{\rho v_r \Theta(j_\mathrm{ISCO}-j)d\cos\theta}, \label{M_fall_BH}\\
   & \dot{M}_\mathrm{fall,disk} = 2\pi r_\mathrm{in}^2\int_{-1}^1{\rho v_r \Theta(j-j_\mathrm{ISCO})d\cos\theta},\label{M_fall_disk}\\
    &\dot{J}_\mathrm{fall,BH}\;\;\, = 2\pi r_\mathrm{in}^2\int_{-1}^1{\rho j v_r \Theta(j_\mathrm{ISCO}-j)d\cos\theta},\\
   & \dot{J}_\mathrm{fall,disk} \;\,= 2\pi r_\mathrm{in}^2\int_{-1}^1{\rho j v_r \Theta(j-j_\mathrm{ISCO})d\cos\theta} \label{J_fall_disk},
\end{align}
where $j(r,\theta)$ is the specific angular momentum and $\Theta(x)$ is the Heaviside step function. 

The mass and angular momentum transfer between the disk and the BH, $\dot{M}_\mathrm{acc}$ and $\dot{J}_\mathrm{acc}$, are estimated as:
\begin{align}
    \dot{M}_\mathrm{acc} &= \frac{M_\mathrm{disk}}{t_{\mathrm{acc}}},\label{M_BH_accretion}\\
    \dot{J}_\mathrm{acc} &= j_{\mathrm{ISCO}}\dot{M}_\mathrm{acc} \label{J_BH_accretion},
\end{align}
with $t_{\mathrm{acc}}$ the accretion time scale which is a free parameter in our models. 
Similarly, the contribution of the disk wind is evaluated as follows:
\begin{align}
    \dot{M}_\mathrm{wind} & = \frac{M_\mathrm{disk}}{t_\mathrm{w}},  \label{M_dot_wind}\\
    \dot{J}_\mathrm{wind} & = j_\mathrm{disk}\dot{M}_\mathrm{wind},  \label{J_dot_wind}
\end{align}
where $t_\mathrm{w}$ is the wind time scale and $j_\mathrm{disk}=J_\mathrm{disk}/M_\mathrm{disk}$  the specific angular momentum of the disk.

Viscosity-driven mechanism is one of the possible mechanisms of the disk outflow. In this scenario, the magnetorotational instability results in the turbulent state in the disk, which acts as the effective viscosity~\citep{Balbus1991a, Balbus:1998ja}. The viscous heating in the disk then drives the wind. There may be another origin of the viscosity: in the surface region of the disk, there is a velocity shear between disk matter and infalling envelope. This may induce the Kelvin-Helmholtz instability, which enhances the magnetic fields leading to the development
of turbulence and dissipating the kinetic energy of the
infalling matter~\citep{Obergaulinger2010jun}. The wind time scale may be different for the different origin of viscosity. There may also be other mechanisms for launching the wind from the disk. For example, the magnetocentrifugal force by large scale magnetic fields can also work to launch the outflow~\citep{Blandford1982jun}. We, therefore, set $t_\mathrm{w}$ as a free parameter not to specify the mechanism for the wind and to investigate more general central engines.

\subsection{Inner boundary condition}\label{sec:Inner_Boundary_Cond}
Until the accretion disk forms, we apply an outflow condition at the inner boundary, thus allowing the material to inflow toward the central engine for $r < r_\mathrm{in}$.
Once the disk is formed, i.e. $M_\mathrm{disk} > 0$, the wind outflow is injected from the inner boundary with the rate of equation~\eqref{M_dot_wind} within an half opening angle $\theta_\mathrm{w}$.
\begin{figure}
	\centering
	\includegraphics [width=0.48\textwidth]{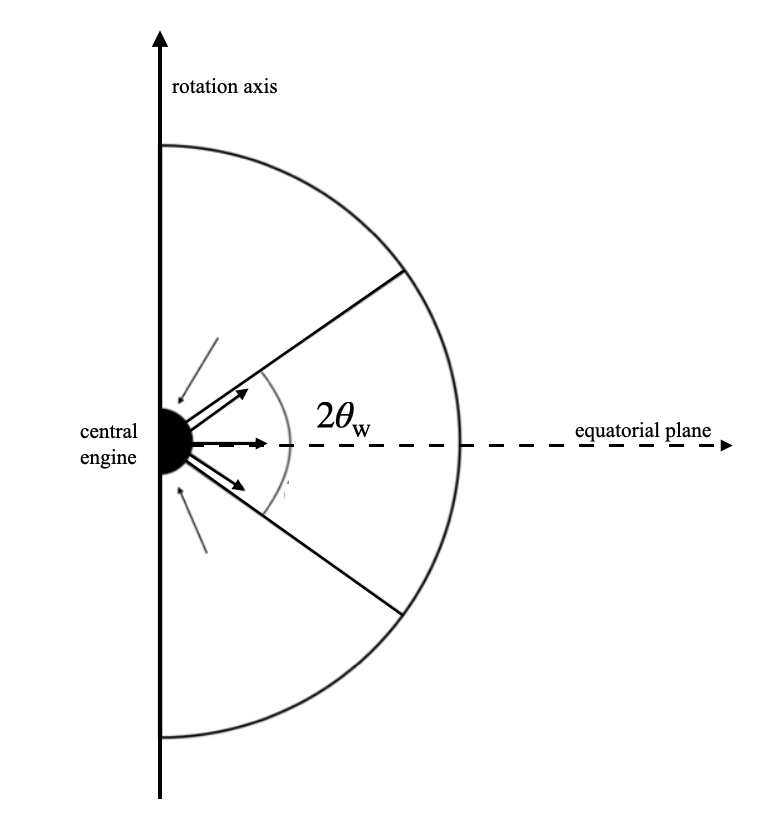}
	\caption{Schematic picture of the explosion in the collapsar scenario. $2\theta_\mathrm{w}$ represents the angle for which we allow the wind outflow. Outside of this angle, the matter is only allowed to infall towards the central engine. The figure also shows the rotation axis and the equatorial plane.}
    \label{fig:scheme_star}
\end{figure}

In Fig.\ref{fig:scheme_star} we present the geometry used in our simulations. The outflow half opening angle at the inner boundary and its orientation can be freely chosen in our code. For this work we set $\theta_\mathrm{w}=\pi/4$ and we direct it along the equatorial plane. 
Within the opening angle of $2\theta_\mathrm{w}$, the wind density is set such that the following relation is satisfied:

\begin{align}
        \dot{M}_\mathrm{wind} = 2\pi r^2_\mathrm{in}\int_{\cos{\theta_1^*}}^{\cos{\theta_2^*}}{\rho_\mathrm{w} v_{\mathrm{w}} d(-\cos\theta)}, \label{rho_from_M_wind}
%        = \frac{f_w}{t_{vis}M_{disk}}
\end{align}

where $\theta_1^*$ and $\theta_2^*$ are the angles of the edges of $2\theta_\mathrm{w}$, and $v_\mathrm{w}$ is the radial velocity of the wind.\footnote{In our method of wind injection, actual mass and energy fluxes at the inner boundary are determined by the numerical fluxes with solving Riemann problem. Therefore, the computed mass flux at the boundary can be smaller than that expected in equation~\eqref{rho_from_M_wind} if, for example, the infalling matter has a larger ram pressure than that of the matter set in ghost cells.}
In this work we decided to describe the outflow density $\rho_\mathrm{w}$ using a parabolic density profile defined as: 
\begin{align}
        \rho_\mathrm{w} = \rho_0(\zeta \cos^2\theta +1), \label{density_profile}
\end{align}

with $\rho_0$ being derived from the integration of equation~\eqref{rho_from_M_wind}.
%integrating the equation~\eqref{rho_from_M_wind}.
In equation~\eqref{density_profile}, we set the parameter ~$\zeta=-1/\cos^2(\pi/2-\theta_\mathrm{w})$ so that the density is zero at the edges of the opening angle (i.e. at $\theta_1=\pi/4$ and $\theta_2=3\pi/4$),
and reaching  maximum value $\rho_0$ at $\theta=\pi/2$, i.e. along the equatorial direction. We define the energy of the disk wind at the inner boundary having as a fraction of the energy related to the disk escape velocity $v_\mathrm{esc}$:

\begin{align}
        \frac{1}{2}v_{\mathrm{w}}^2 +f_\mathrm{therm} \frac{1}{2}v_{\mathrm{w}}^2+\Phi = \xi^2 \frac{1}{2}v_\mathrm{esc}^2 \label{e_inj_eqs},
\end{align}

where $v_\mathrm{esc}$ is:

\begin{equation}\label{v_esc}
    v_\mathrm{esc}=\sqrt{\frac{2G M_\mathrm{BH}}{r_\mathrm{disk}}} ,
\end{equation}
with the disk radius $r_\mathrm{disk}$ defined as:
\begin{equation}\label{r_disk_definition}
    r_\mathrm{disk}= j_\mathrm{disk}^2/{GM_\mathrm{BH}}.
\end{equation}

In equation~\eqref{e_inj_eqs} the internal energy of the wind is given by $e_{\mathrm{int,w}}/\rho_\mathrm{w}=(1/2)f_\mathrm{therm} v_{\mathrm{w}}^2$ .  $f_\mathrm{therm}$ is a free parameter in our simulations, measuring the fraction of the wind kinetic energy assumed to be corresponding to its internal energy.  $\xi$ is a fudge factor used to represent the uncertainties coming from the lack of knowledge of the precise disk structure \citep{Hayakawa_2018}, and it is a free parameter in our simulations. 
The pressure of the outflow is computed using the tabulated EOS with the density $\rho_\mathrm{w}$ and the internal energy $e_{\mathrm{int,w}}$ as input parameters. 
equation~\eqref{e_inj_eqs} indicates that the asymptotic velocity of the injected matter, the velocity of the matter at infinite distance in the case the total specific energy $(1/2)v^2+e_\mathrm{int}/\rho+\Phi$ is conserved, is $\xi v_\mathrm{esc}$. 

A part of the injected matter could fall back to the central engine, affecting the disk mass. This is the case when the ram pressure of the matter at the inner boundary is larger than that of the injected matter. To avoid the recycling of the injected matter, in our study we set the angular momentum in the ghost cells to zero. In this way we do not allow the injected matter to fall back to the disk, but only to the BH.

For $\theta<\theta^*_1$ or $\theta>\theta^*_2$, the boundary condition is set to prevent the matter from inflowing from the central engine to the computational domain. To achieve that we set zero fluxes (reflecting boundary condition) if the radial velocity in a first active cell is positive, while we allow the mass infall to the central engine if it is negative.

\subsection{The equation of state}
The thermodynamical properties of the star are described by a tabulated equation of state (EOS)
that includes the ion, the radiation, the electrons, and $e^-$-- $e^+$ pair. In this work we use an oxygen-based EOS, i.e an EOS using oxygen as the only component of the ion (i.e. $Y_{e}=0.5$), resulting in a $^{16}$O mass fraction of 1. This decision is made considering the composition of our progenitor model dominated by oxygen outside the iron core (see \citealt{Aguilera-Dena_2020}). 

\subsection{Diagnostics}\label{sec:Diagnostic}
In the following subsection, we describe the method used to calculate the properties of  the ejecta and injected matter. 

In our simulations, we define the ejecta mass $M_\mathrm{ej}$ as the sum of unbound matter mass.
The explosion energy $E_{\mathrm{expl}}$ is the energy carried by the unbound matter. Several criteria are used to define the unbound matter in hydrodynamic simulations (citation here). For our study we chose to use the Bernoulli criterion which takes into account the thermal effect on the matter and the effects of the gravitational potential, and is defined as follows:

\begin{align}
     B := \frac{e_\mathrm{int}+e_\mathrm{kin}+P}{\rho}+\Phi>0. \label{Bernoulli_criterion}
\end{align}
Using the Bernoulli criterion we track the evolution of the ejecta mass\footnote{Using the Bernoulli criterion to compute the ejecta mass, we are actually defining it as the unbound mass.} and energy at every time step by integrating the equations:
\begin{align}
M_\mathrm{ej} = r_\mathrm{out}^2 \int_0^t \int_{B>0,v_r>0} \rho v_r d\Omega dt + \int_{B>0,v_r>0} \rho d^3x ,
\end{align}
\begin{align}
E_\mathrm{expl} & =  r_\mathrm{out}^2 \int_0^t \int_{B>0,v_r>0} \rho B v_r d\Omega dt +\int_{B>0,v_r>0} (e_\mathrm{t} +\rho\Phi) d^3x .
\end{align}

The injected mass $M_\mathrm{inj}$ represents the matter coming from the central engine with a positive mass flux at the inner boundary $r_\mathrm{in}$. It is defined as:
\begin{align} \label{M_injected_def}
    M_\mathrm{inj} = r_\mathrm{in}^2 \int_0^t \int_{B>0,v_r>0} \rho v_r d\Omega dt.
\end{align}

We consider the injected energy $E_\mathrm{inj}$ as the energy carried by $M_\mathrm{inj}$ with positive binding energy.% but it also takes into account the effect of the gravitational potential, therefore it is defined as the energy of that part of $M_\mathrm{inj}$ with positive binding energy. We then use again the Bernoulli criterion to evaluate it.
We, then, compute the injected energy $E_\mathrm{inj}$ applying the Bernoulli criterion as follows:
\begin{align}
E_\mathrm{inj}  =  r_\mathrm{in}^2 \int_0^t \int_{B>0,v_r>0} \rho B v_r d\Omega dt.
\end{align}

\subsection{Parameters and initial condition}\label{parameter_initial_C}
In this work, as mentioned above, we consider the scenario of a failed CCSN and we assume that the mass of the innermost region ($1.28\,M_\odot$) of our progenitor corresponds to the initial mass of the BH.  In our model, the beginning of the disk formation is when the condition to launch the wind from the inner boundary is met for the first time. For our simulations, the density and velocity structure of the wind are fixed as explained in Sec.~\ref{sec:Inner_Boundary_Cond}. We then use four more free parameters, which are the wind time scale $t_\mathrm{w}$, the ratio between the accretion and wind time scales $t_\mathrm{acc}/t_\mathrm{w}$, the ratio between the radial velocity of the outflow and the escape velocity $\xi$ , and $f_{\mathrm{therm}}$ which measures the fraction of the wind kinetic energy assumed to be corresponding to its internal energy (see equation~\eqref{e_inj_eqs}). In this work, we fix the direction of the outflow, its opening angle, and the density profile as we want to investigate the parameter space of the other quantities.
Specifically, we set the wind along the equatorial plane using an half-opening angle $\theta_\mathrm{w}$ of $\pi/4$, and the density profile $\rho_\mathrm{w}$ described in equation~\eqref{density_profile}.

Using this setup, we investigate the parameter space for $t_\mathrm{w}$, $t_\mathrm{acc}/t_\mathrm{w}$, $\xi$ and $f_{\mathrm{therm}}$. 
We sample $t_\mathrm{w}$ in a wide interval, $(0.1, 1, 3.16, 10)$ s, using also more extreme values like 0.1 or 10~s (usually $t_\mathrm{w}$ is few seconds as shown by~\citealt{Wang_2023}) to survey a parameter space as large as possible and to analyse the condition to reach the energies and the amount of $^{56}$Ni produced in high-energy SNe.
$t_\mathrm{acc}$ is set through the ratio $t_\mathrm{acc}/t_\mathrm{w}$ and the value of $t_\mathrm{w}$. We vary  $t_\mathrm{acc}/t_\mathrm{w}$ in the interval (1,3.16,10, $\infty$). The accretion time scale controls the accretion rate onto the BH from the disk,
therefore it allows to track the dynamics of the central engine (see also \citealt{Kumar_Narayan_method}).
If $t_\mathrm{acc}$ has small values so that it is shorter than the infalling time scale of the envelope (which is given by $\dot{M}_\mathrm{fall,disk}/M_\mathrm{disk}$), then the accretion rate onto the BH tracks the rate at which mass is falling onto the disk. 
On the contrary, for longer $t_\mathrm{acc}$ up to the extreme case of $t_\mathrm{acc}=\infty$ the mass infall onto the disk dominates. Varying $t_\mathrm{acc}/t_\mathrm{w}$ from 1 to $\infty$ allows us to investigate the effect of these two very different scenarios on the explosion and on the $^{56}$Ni production.
We assume the wind time scale and accretion time scale to be constant throughout the explosion in order to model the central engine as simple as possible.

In our simulations we use $\xi^2=(0.1, 0.3)$ following the approach of \citet{Hayakawa_2018} who used $\xi^2=0.1$ in their work. We increase it because of our interest in the high energy explosions. 

Finally, we set $f_\mathrm{therm}$ as $(0.1, 0.01)$ following the typical values of the wind internal energy in the literature (as in \citealt{Hayakawa_2018}). In our work, we, then, test several combinations of these parameters.

\subsection{Tracer particles and nucleosynthesis}\label{sec:Nucleosynthesis}
To obtain thermodynamic histories of the ejecta, we use tracer particles following the method described in \cite{Fujibayashi2023jan}. 
In this method, the evolution of tracer particles is followed backward in time. Hence, they are placed every time interval $\Delta t$ from the end of the simulation at radius $r=r_\mathrm{out}$ in the range of $0\leq \theta \leq \pi$.
The mass of each particle is defined as $\Delta m = \rho v_r {r_\mathrm{ext}}^2\Delta \Omega \Delta t$, where $\Delta \Omega$ is the solid angle element. The time interval $\Delta t$  is defined as  as $\Delta t := r_\mathrm{out}\Delta\theta/\langle v_r\rangle$, where $\Delta \theta$ is the interval of the polar angle and $\langle v_r\rangle$ is the average radial velocity of the ejecta at $r=r_\mathrm{out}$. This formulation of $\Delta t$ ensures an optimal distribution of tracer particles in time. 

This method is also utilized to judge whether a given fluid element is the injected matter from the inner boundary or the one that originates from the stellar envelope: A particle is tagged as an injected matter if it crosses the inner boundary during the back-tracing. On the other hand, if a particle stays inside the computational domain until the initial snapshot, it is an ejecta component originating from the stellar envelope.

A disadvantage of using a post-process particle tracing method is that the accuracy of the thermodynamics histories is limited by the frequency of the output (see, e.g., \citealt{Sieverding2023jun}). For most of the model, the time interval of the output is $\sim$ 70~ms. To check the systematic error that stems from this limitation, we performed a set of simulations using a higher time resolution of $\Delta t \sim$ 10~ms for selected models, and performed particle tracing using this time interval. This convergence study is presented in Appendix~\ref{Appendix:output_dependence_particle_tracing}.

For tracer particles originating from the stellar envelope, the full thermodynamical history is available. To obtain the nucleosynthetic yield based on the density and temperature evolution along such particles, we perform nucleosynthesis calculation with the open-source code \texttt{torch} \citep{Timmes2000jul} using 200 isotopes. The initial composition of the calculation is set by the stellar composition at the initial position of each tracer particle. Due to the lack of knowledge about the thermodynamical history of the injected matter (coming from the inner boundary), we perform nucleosynthesis calculation only for the stellar envelope.

Since we carry out the nucleosynthesis calculation without evolving the stellar composition from that of the original star (hence assuming always symmetric matter) and without taking into account neutrino interactions, the $^{56}$Ni mass evaluated might be slightly higher than in reality. Yet, the nucleosynthesis calculation is performed only for those particles which are inside the computational domain for the whole simulation (i.e. at $r>r_\mathrm{in}=10^8$ cm) where the neutrino interaction is not significant and hence not expected to strongly affect the $^{56}$Ni production.

\section{Results}\label{sec:Results}

In this section, we will first describe the hydrodynamical evolution of the supernova explosion using one of our simulations as the characteristic model (subsection~\ref{sec:hydrodynamics}). Then, in the subsection ~\ref{sec:model_comparison}, we will compare the results of some observables between our different models, using this to analyse the effect of  $t_\mathrm{w}$, $t_\mathrm{acc}/t_\mathrm{w}$, $\xi$ and $f_{\mathrm{therm}}$  on the explosion and $^{56}$Ni production. 

\begin{figure}
	\centering
	\includegraphics [width=0.48\textwidth]{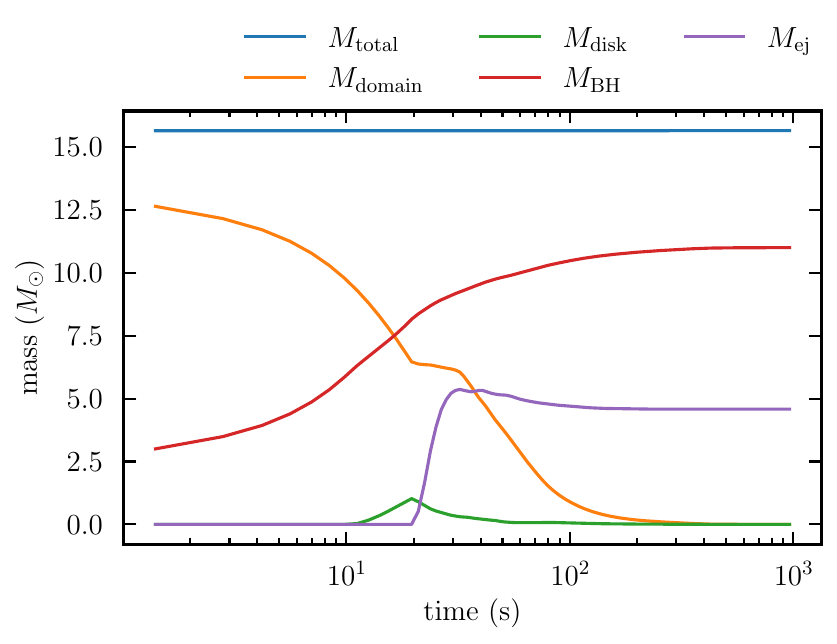}
	\caption{Masses evolution for the characteristic model , M20\_10\_1\_0.3\_0.1. In cyan the total mass, in orange the mass enclosed in our computational domain, in green the disk mass, in red the BH mass and in purple the ejecta mass.  $M_\mathrm{total}$ is plotted to show the conservation of mass throughout the simulation. 
	}
    \label{fig:mass_components_charact_model}
\end{figure}

\begin{figure}
	\centering
	\includegraphics [width=0.48\textwidth]{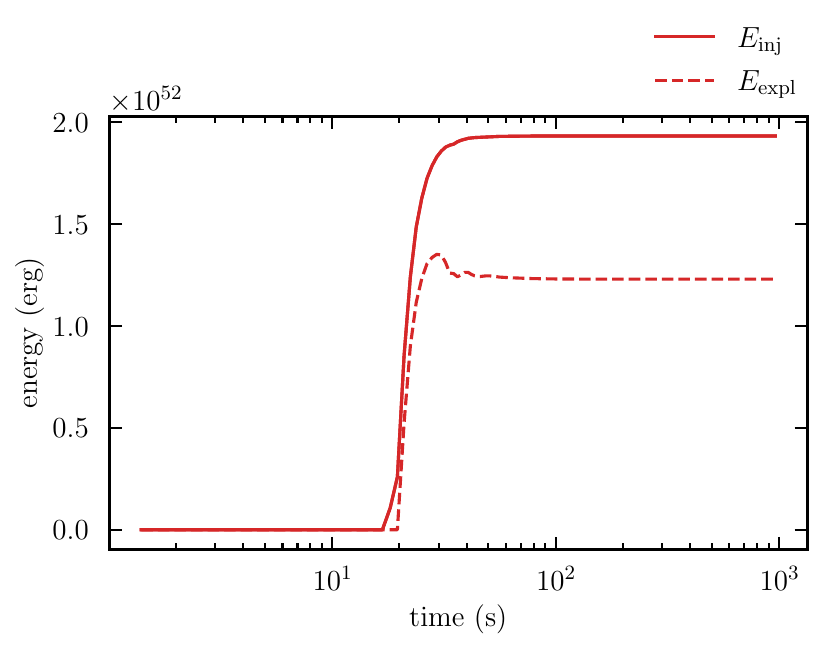}
    	\caption{Injected (solid line) and explosion energy (dashed line) evolution for the characteristic model, M20\_10\_1\_0.3\_0.1.}
       \label{fig:E_inj_E_expl_charact_model}
\end{figure}

\begin{table*}
\centering
\caption{Model description and key results. From left to right, the columns contain wind time scale, the ratio of the accretion and wind time scales, the squared ratio of the asymptotic velocity of injected matter to escape velocity of the disk, the internal to kinetic energy ratio of injected matter, cumulative injected energy, ejecta mass, explosion energy, average ejecta velocity, the mass of ejecta component that is originated from the computational domain and experienced temperature higher than 5\,GK, the mass of the $^{56}$Ni synthesized, the mass of ejecta component originated from the injected matter. The results are for an output frequency of  70~ms.
}
\begin{tabular}{lccccccccccc}
\hline\hline
model &$t_\mathrm{w}$ & $t_\mathrm{acc}/t_\mathrm{w}$ & $\xi^2$ & $f_\mathrm{therm}$ & $E_\mathrm{inj}$ & $M_\mathrm{ej}$ & $E_\mathrm{expl}$ & $v_\mathrm{ej}$ & $M_\mathrm{ej,>5GK}^\mathrm{stellar}$ & $M_\mathrm{ej,Ni}^\mathrm{stellar}$ & $M_\mathrm{ej}^\mathrm{inj}$ \\ %& $M_\mathrm{Ni}^\mathrm{inj}$ \\
&(s) & & & & ($10^{51}$\,erg) & ($M_\odot$) & ($10^{51}$\,erg) & ($10^3$\,km/s) & ($M_\odot$) & ($M_\odot$) & ($M_\odot$) \\ %& ($M_\odot$) \\ 
\hline
M20\_0.1\_1\_0.1\_0.10 & 0.1 &        1 & 0.1 & 0.10 & $<0.01$ &  0.60 & 0.049 &   2.9 &           0.014 &           0.016 &          0.0077 \\
M20\_0.1\_3.16\_0.1\_0.10 &  0.1 &     3.16 & 0.1 & 0.10 &    0.36 &  0.64 & 0.057 &   3.0 &          0.0070 &          0.0069 &           0.015 \\
M20\_0.1\_10\_0.1\_0.10 &  0.1 &       10 & 0.1 & 0.10 &    0.80 &  0.77 & 0.057 &   2.7 &           0.016 &           0.028 &           0.015 \\
M20\_0.1\_inf\_0.1\_0.10 &  0.1 & $\infty$ & 0.1 & 0.10 &     1.1 &  0.78 & 0.051 &   2.6 &           0.024 &           0.034 &           0.012 \\
M20\_1\_1\_0.1\_0.10 &    1 &        1 & 0.1 & 0.10 &    0.12 &  0.63 & 0.073 &   3.4 &          0.0089 &           0.013 &           0.032 \\
M20\_1\_3.16\_0.1\_0.10 &     1 &     3.16 & 0.1 & 0.10 &     3.3 &  0.65 &  0.24 &   2.8 &           0.040 &           0.048 &           0.013 \\
M20\_1\_10\_0.1\_0.10 &     1 &       10 & 0.1 & 0.10 &     4.8 &  0.77 & 0.077 &   3.2 &           0.025 &           0.028 &           0.022 \\
M20\_1\_inf\_0.1\_0.10 &     1 & $\infty$ & 0.1 & 0.10 &     4.3 &   1.3 &  0.24 &   4.3 &          0.0054 &          0.0083 &           0.028 \\
M20\_3.16\_1\_0.1\_0.10 &  3.16 &        1 & 0.1 & 0.10 &     1.1 &  0.64 & 0.049 &   2.8 &          0.0088 &           0.020 &          0.0079 \\
M20\_3.16\_3.16\_0.1\_0.10 &  3.16 &     3.16 & 0.1 & 0.10 &     7.1 &   3.4 &   3.0 &   9.4 &           0.036 &           0.035 &             1.1 \\
M20\_3.16\_10\_0.1\_0.10 &  3.16 &       10 & 0.1 & 0.10 &     8.6 &   4.2 &   4.4 &    10 &           0.048 &           0.046 &             1.4 \\
M20\_3.16\_inf\_0.1\_0.10 &  3.16 & $\infty$ & 0.1 & 0.10 &     9.2 &   4.2 &   4.6 &    10 &           0.050 &           0.049 &             1.2 \\
M20\_10\_1\_0.1\_0.10 &    10 &        1 & 0.1 & 0.10 &     1.8 &  0.71 & 0.062 &   3.0 &          0.0076 &           0.011 &           0.019 \\
M20\_10\_3.16\_0.1\_0.10 &    10 &     3.16 & 0.1 & 0.10 &      12 &   4.4 &   7.7 &    13 &   0.053  &   0.039  &       1.2  \\

M20\_10\_10\_0.1\_0.10 &    10 &       10 & 0.1 & 0.10 &      15 &   5.3 &    11 &    14 &   0.061  &   0.052 &       1.9  \\

M20\_10\_inf\_0.1\_0.10 &    10 & $\infty$ & 0.1 & 0.10 &      17 &   5.9 &    12 &    14 &   0.064  &   0.061  &       2.0 \\

M20\_0.1\_1\_0.3\_0.10 &   0.1 &        1 & 0.3 & 0.10 & $<0.01$ &  0.72 & 0.055 &   2.6 &          0.0066 &           0.010 &           0.018 \\
M20\_0.1\_3.16\_0.3\_0.10 &   0.1 &     3.16 & 0.3 & 0.10 &     1.2 &  0.73 & 0.067 &   3.0 &           0.020 &           0.025 &           0.022 \\
M20\_0.1\_10\_0.3\_0.10 &   0.1 &       10 & 0.3 & 0.10 &     2.2 &  0.87 & 0.070 &   2.8 &           0.024 &           0.035 &           0.016 \\
M20\_0.1\_inf\_0.3\_0.10 &   0.1 & $\infty$ & 0.3 & 0.10 &     2.6 &  0.87 & 0.079 &   3.0 &           0.037 &           0.056 &           0.014 \\
M20\_1\_1\_0.3\_0.10 &  1 &        1 & 0.3 & 0.10 &     3.1 &  0.75 & 0.088 &   3.5 &           0.013 &           0.024 &           0.022 \\
M20\_1\_3.16\_0.3\_0.10  &  1 &     3.16 & 0.3 & 0.10 &     6.6 &   3.7 &   1.6 &   6.6 &            0.39 &            0.35 &             1.3 \\
M20\_1\_10\_0.3\_0.10  &  1 &       10 & 0.3 & 0.10 &     7.7 &   4.2 &   3.6 &   9.3 &           0.037 &           0.096 &             1.1 \\
M20\_1\_inf\_0.3\_0.10   & 1 & $\infty$ & 0.3 & 0.10 &     7.7 &   4.0 &   3.5 &   9.4 &           0.049 &           0.058 &             1.1 \\
M20\_3.16\_1\_0.3\_0.10 &3.16 &        1 & 0.3 & 0.10 &      10 &   3.0 &   3.0 &    10 &           0.048 &           0.044 &            0.39 \\
M20\_3.16\_3.16\_0.3\_0.10 &3.16 &     3.16 & 0.3 & 0.10 &      14 &   5.4 &   8.0 &    12 &   0.051  &   0.062 &     0.80  \\

M20\_3.16\_10\_0.3\_0.10 &3.16 &       10 & 0.3 & 0.10 &      15 &   5.8 &   9.2 &    13 &   0.039 &   0.065 &     0.69 \\

M20\_3.16\_inf\_0.3\_0.10 &3.16 & $\infty$ & 0.3 & 0.10 &      16 &   6.0 &    11 &    13 &   0.044 &   0.072 &       1.0 \\

M20\_10\_1\_0.3\_0.10 &  10 &        1 & 0.3 & 0.10 &      19 &   4.6 &    12 &    16 &   0.041 &   0.052 &     0.59\\

M20\_10\_3.16\_0.3\_0.10 &  10 &     3.16 & 0.3 & 0.10 &      32 &   5.9 &    25 &    21 &   0.054  &   0.079  &     0.78 \\

M20\_10\_10\_0.3\_0.10 &  10 &       10 & 0.3 & 0.10 &      38 &   6.6 &    32 &    22 &     0.13 &     0.14 &       1.4 \\

M20\_10\_inf\_0.3\_0.10 &  10 & $\infty$ & 0.3 & 0.10 &      40 &   6.8 &    34 &    23 &    0.076  &   0.096  &       1.4  \\

M20\_0.1\_1\_0.1\_0.01 &  0.1 &        1 & 0.1 & 0.01 & $<0.01$ &  0.66 & 0.053 &   2.8 &           0.020 &           0.023 &           0.019 \\
M20\_0.1\_3.16\_0.1\_0.01 &  0.1 &     3.16 & 0.1 & 0.01 &    0.36 &  0.68 & 0.060 &   3.0 &          0.0057 &          0.0082 &           0.016 \\
M20\_0.1\_10\_0.1\_0.01 &  0.1 &       10 & 0.1 & 0.01 &    0.79 &  0.73 & 0.073 &   3.2 &           0.036 &           0.061 &           0.012 \\
M20\_0.1\_inf\_0.1\_0.01 &  0.1 & $\infty$ & 0.1 & 0.01 &     1.1 &  0.81 & 0.076 &   3.1 &           0.017 &           0.023 &          0.013 \\
M20\_1\_1\_0.1\_0.01 &    1 &        1 & 0.1 & 0.01 &    0.20 &  0.61 & 0.044 &   2.7 &          0.0063 &           0.013 &          0.0073 \\
M20\_1\_3.16\_0.1\_0.01 &    1 &     3.16 & 0.1 & 0.01 &     3.0 &  0.88 & 0.088 &   3.2 &            0.18 &            0.21 &          0.0029 \\
M20\_1\_10\_0.1\_0.01 &    1 &       10 & 0.1 & 0.01 &     3.6 &  0.92 & 0.077 &   2.9 &           0.095 &            0.13 &          0.0055 \\
M20\_1\_inf\_0.1\_0.01 &    1 & $\infty$ & 0.1 & 0.01 &     3.8 &  0.98 &  0.25 &   5.0 &         0.00012 &          0.0097 &           0.059 \\
M20\_3.16\_1\_0.1\_0.01 & 3.16 &        1 & 0.1 & 0.01 &     1.4 &  0.58 & 0.053 &   3.0 & 0.0083  &   0.011  & 0.0090 \\

M20\_3.16\_3.16\_0.1\_0.01 & 3.16 &     3.16 & 0.1 & 0.01 &     7.3 &   2.8 &   2.5 &   9.6 &   0.055  &   0.052 &     0.58  \\

M20\_3.16\_10\_0.1\_0.01 & 3.16 &       10 & 0.1 & 0.01 &     8.6 &   3.9 &   4.5 &    11 &   0.063 &   0.062  &       1.1  \\

M20\_3.16\_inf\_0.1\_0.01 & 3.16 & $\infty$ & 0.1 & 0.01 &     9.5 &   4.3 &   5.0 &    11 &    0.084 &   0.083 &       1.1 \\

M20\_10\_1\_0.1\_0.01 &   10 &        1 & 0.1 & 0.01 &      23 &  0.68 & 0.075 &   3.4 &           0.011 &           0.015 &           0.031 \\
M20\_10\_3.16\_0.1\_0.01 &   10 &     3.16 & 0.1 & 0.01 &      12 &   5.3 &   8.2 &    12 &   0.057 &   0.054  &       1.8 \\

M20\_10\_10\_0.1\_0.01 &  10 &       10 & 0.1 & 0.01 &      14 &   5.7 &   9.9 &    13 &   0.072  &   0.061  &       1.9  \\

M20\_10\_inf\_0.1\_0.01 &   10 & $\infty$ & 0.1 & 0.01 &      17 &   5.9 &    12 &    14 &   0.059 &   0.058  &       2.3  \\

\hline
\end{tabular}
\label{tab:models}
\end{table*}

\subsection{Hydrodynamical evolution}\label{sec:hydrodynamics}
To present the general outline of the evolution of our simulations, we describe the model, M20\_10\_1\_0.3\_0.1, with parameters   $t_\mathrm{w} = 10$~s, $t_\mathrm{acc}/t_\mathrm{w} = 1$, $\xi^2 = 0.3$ and $f_{\mathrm{therm}} = 0.1$ as example. 
In Fig.~\ref{fig:mass_components_charact_model}, we plot the evolution of the total mass, BH mass, disk mass,  mass enclosed in our computational domain, and ejecta mass for this model. In the first $\sim 10$ seconds, the infalling matter accretes only onto the BH because of the small angular momentum of infalling matter.  The disk formation starts after about 10~s, which is the condition to trigger the wind injection  in our simulation.
A part of the disk mass accretes to the BH (equation~\eqref{M_BH_accretion}) and the other part contributes to the wind according to equation \eqref{M_dot_wind}. The injection of the wind does not occur for the first 10-20~s due to the small ram pressure of the injected matter compared to that of the infalling matter. In this model it starts at $t\approx \SI{20}{s}$ and it leads to the increase in the ejecta mass. The disk mass peaks at about $1\,M_\odot$, followed by a decrease because of the rate of the mass accretion into the BH and injection as the wind mass exceeds that of the supply to the disk from the stellar envelope. 
After $\sim$200~s, the mass components reach approximately constant values and $M_\mathrm{BH}$ becomes $\sim 11\, M_\odot$.
The ejecta mass reaches a temporal maximum beyond $5\,M_\odot$ after $\sim32$~s and then converges to $4.6\,M_\odot$.

In Fig.~\ref{fig:E_inj_E_expl_charact_model} we show the time evolution of the injected and explosion energies of the characteristic model, M20\_10\_1\_0.3\_0.1. It shows a wind injection beginning at $\sim 20$~s and lasting $\sim20$~s. Within the first 40 seconds from the start of the simulation the wind has already been almost injected, reaching an energy $E_\mathrm{inj}$ of $\sim 19.33 \times 10^{51}$ erg.
In this model, we see that towards the end of the wind injection period, $E_\mathrm{expl}$ reaches a temporal maximum, before plateauing after the wind injection is finished at $E_\mathrm{expl}\sim12.30 \times10^{51}$ erg.

 \begin{figure*}
\centering
\includegraphics [width=0.489\textwidth]{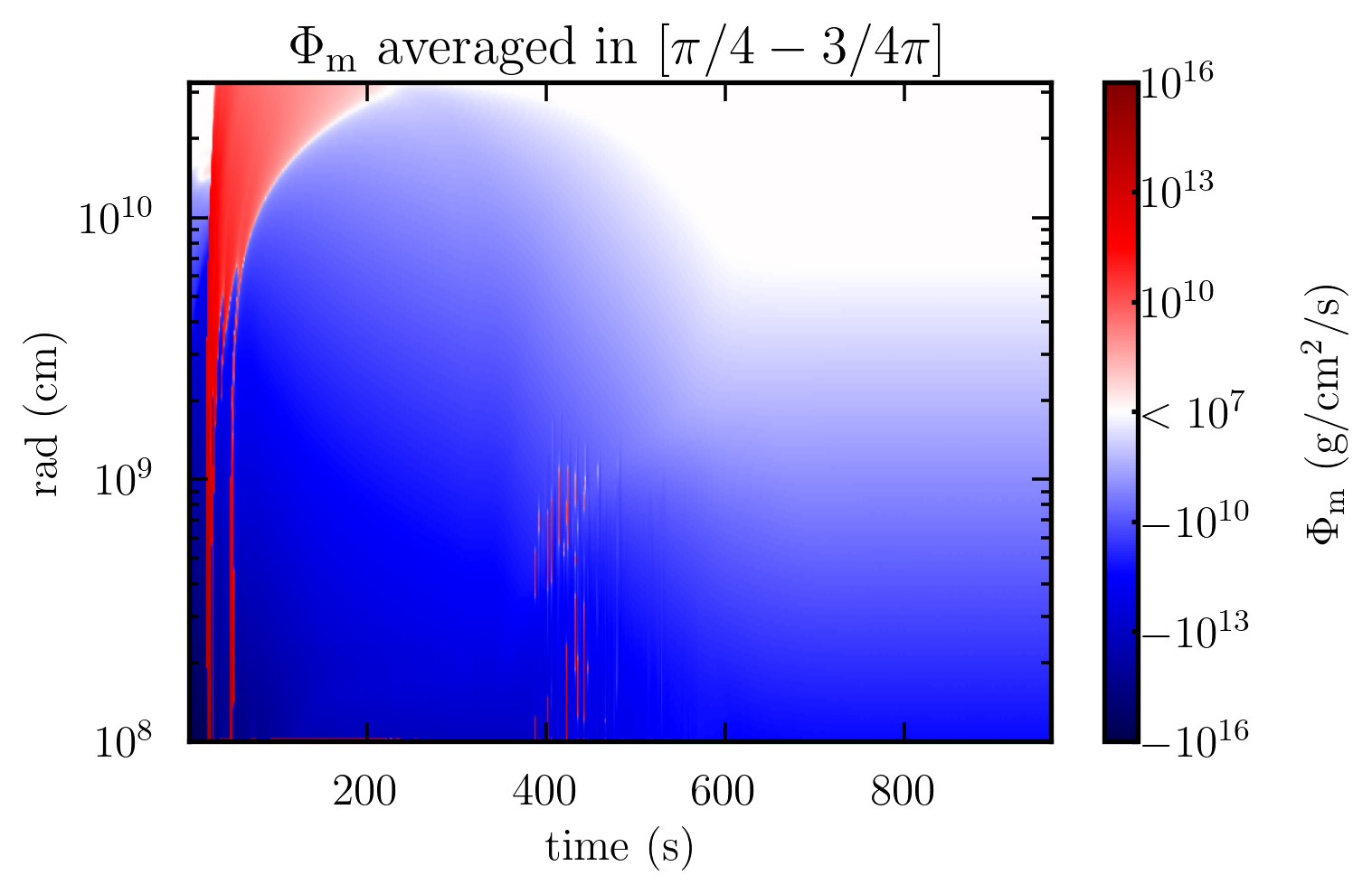}
\includegraphics [width=0.48\textwidth]{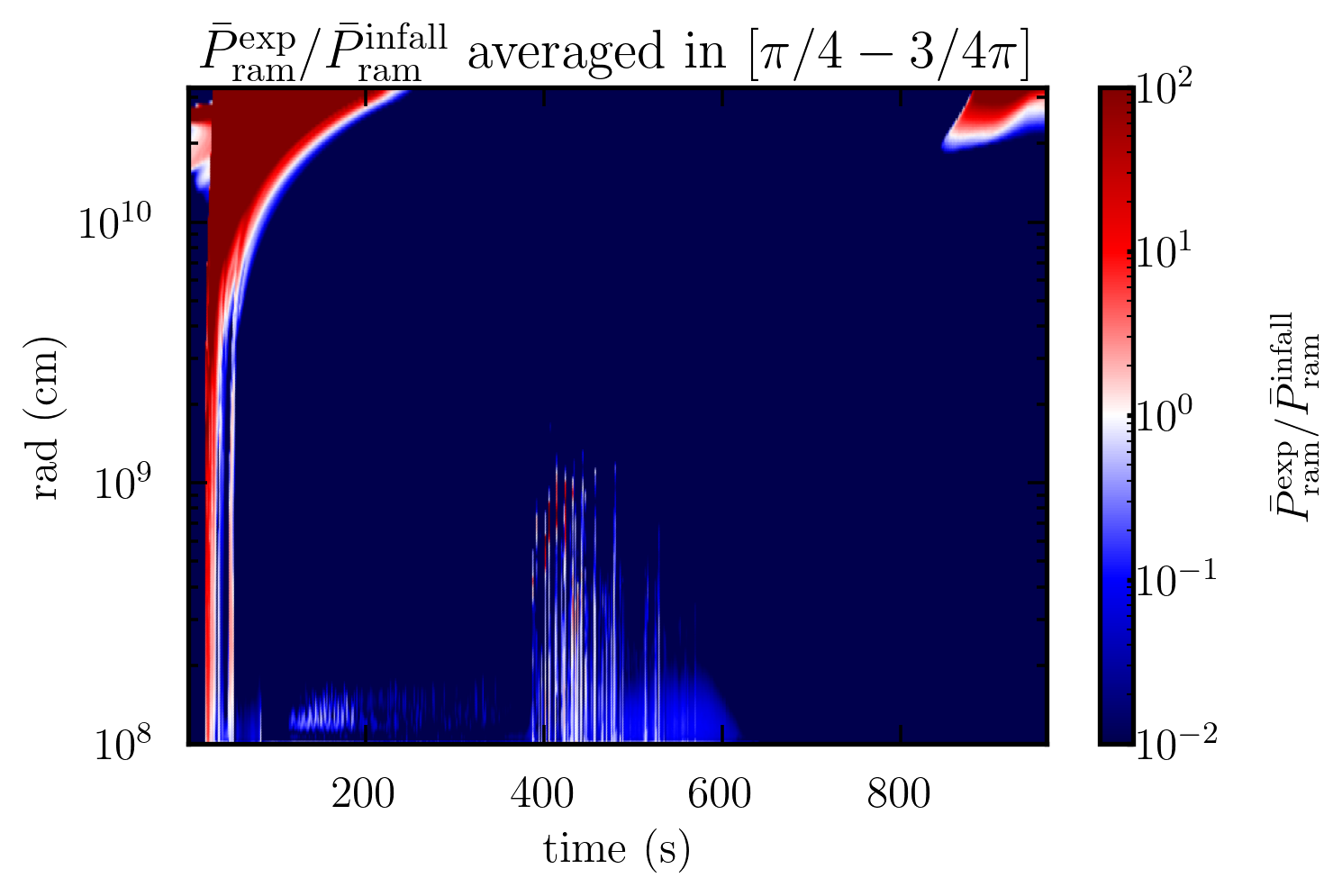}
\caption{Left panel: time evolution of the mass outflow rate $\Phi_\mathrm{m}$ averaged over the injection angle (i.e., in $[\pi/4 - 3/4\pi]$).  Right panel: space-time diagram of the ratio between the ram pressure averaged over the injection angle of the injected matter, $\bar{P}_\mathrm{ram}^\mathrm{exp}$, and of the infalling envelope, $\bar{P}_\mathrm{ram}^\mathrm{infall}$. These plots are obtained for the model M20\_10\_1\_0.3\_0.1.}
\label{fig:flux_m_Pram_spacetime_tw10}
\end{figure*}

\begin{figure}
\centering
	\captionsetup{font=small, labelfont=rm}
	%\subfloat[]
	{\includegraphics [width=0.48\textwidth]{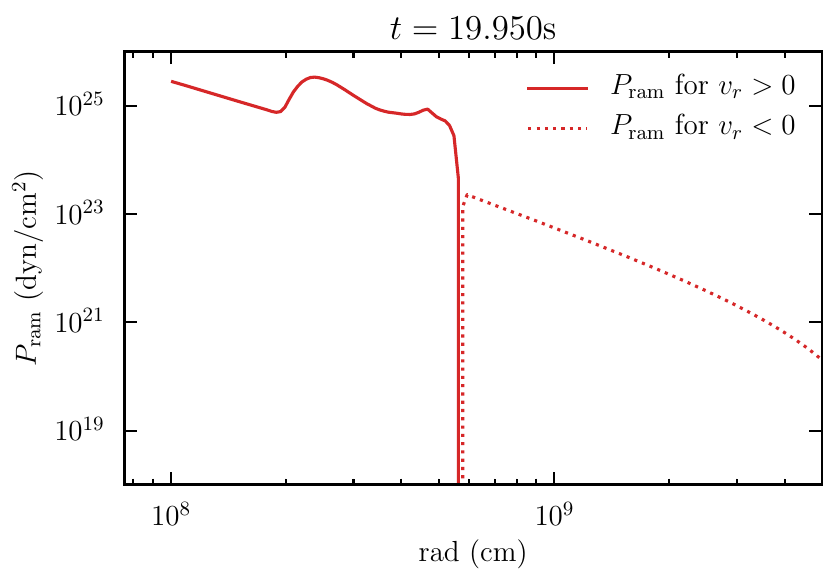}	
		\label{fig:P_ram_charact_model} }\quad
	%\subfloat[]
	{\includegraphics [width=0.48\textwidth]{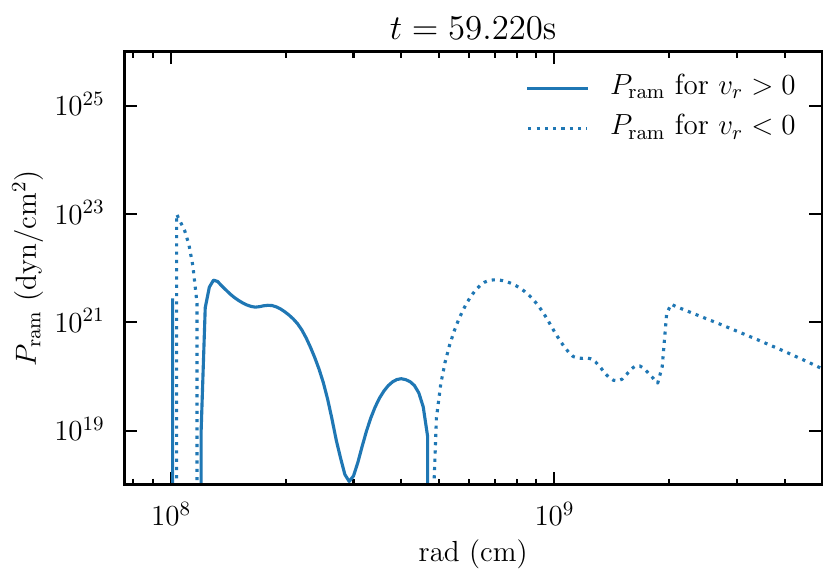}
		\label{fig:P_ram_tw1} }
\caption{Radial distribution along the equator ($\theta=\pi/2$) of $P_\mathrm{ram}$ for the expanding matter (solid line) and the infalling matter (dotted line) up to $r=3\times10^9$ cm, after the wind onset. The distribution is compared between the models  M20\_10\_1\_0.3\_0.1 and  M20\_1\_1\_0.3\_0.1. Respectively, 
the upper panel shows the distribution for  M20\_10\_1\_0.3\_0.1 with high $E_\mathrm{expl}/E_\mathrm{inj}$ and the lower panel displays the radial distribution of P$_{\rm ram}$ for  M20\_1\_1\_0.3\_0.1. The comparison time is chosen such as the front of the outflow is at similar radius for both simulations. 
%They are compared at times for which the front of the outflow is approximately at the same radius.
	}
\label{fig:P_ram_2_models_comparison}
%\label{fig:P_ram_charact_model}
\end{figure}

\begin{figure*}
\centering
\includegraphics [width=0.489\textwidth]{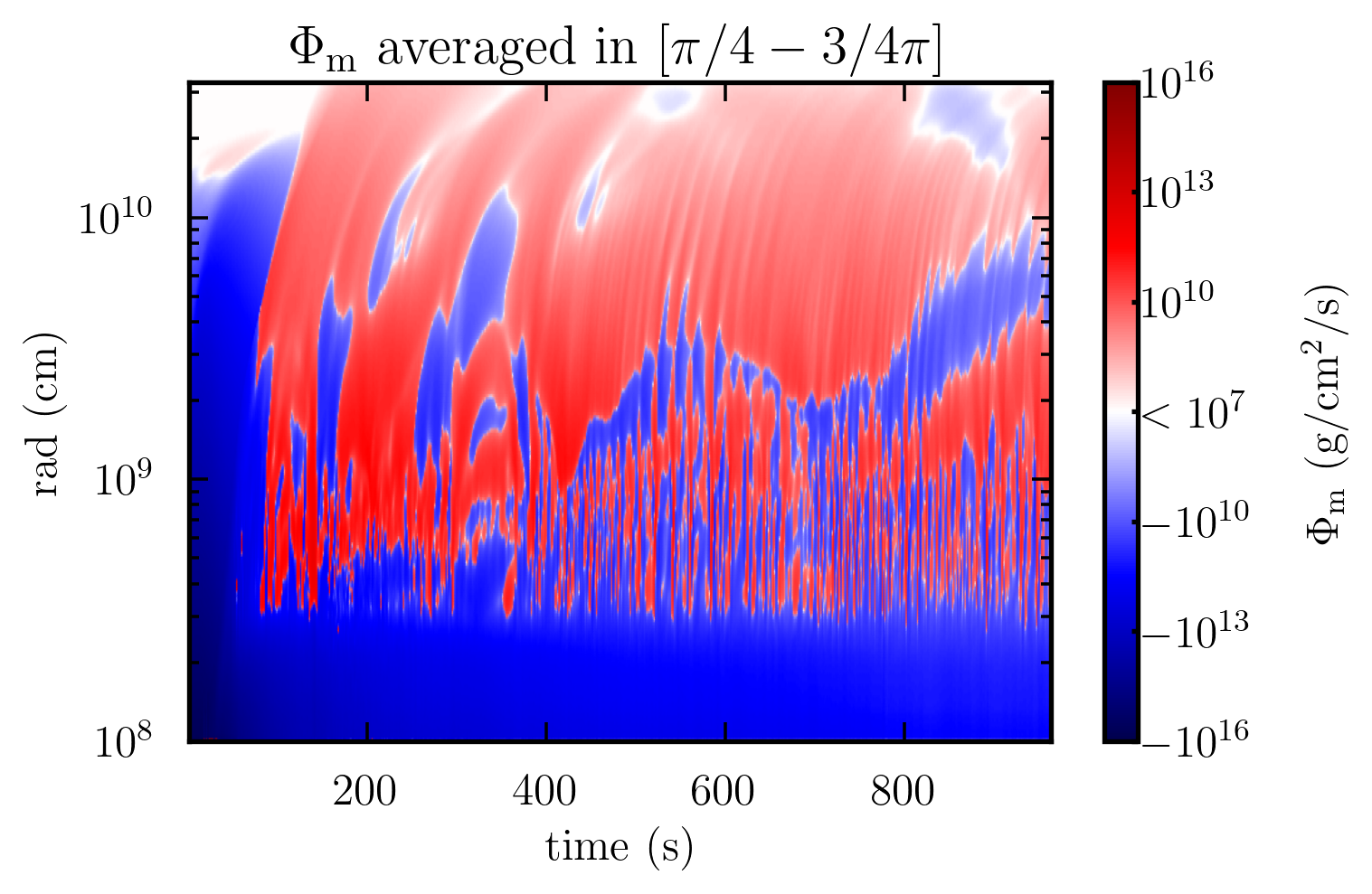}
\includegraphics [width=0.48\textwidth]{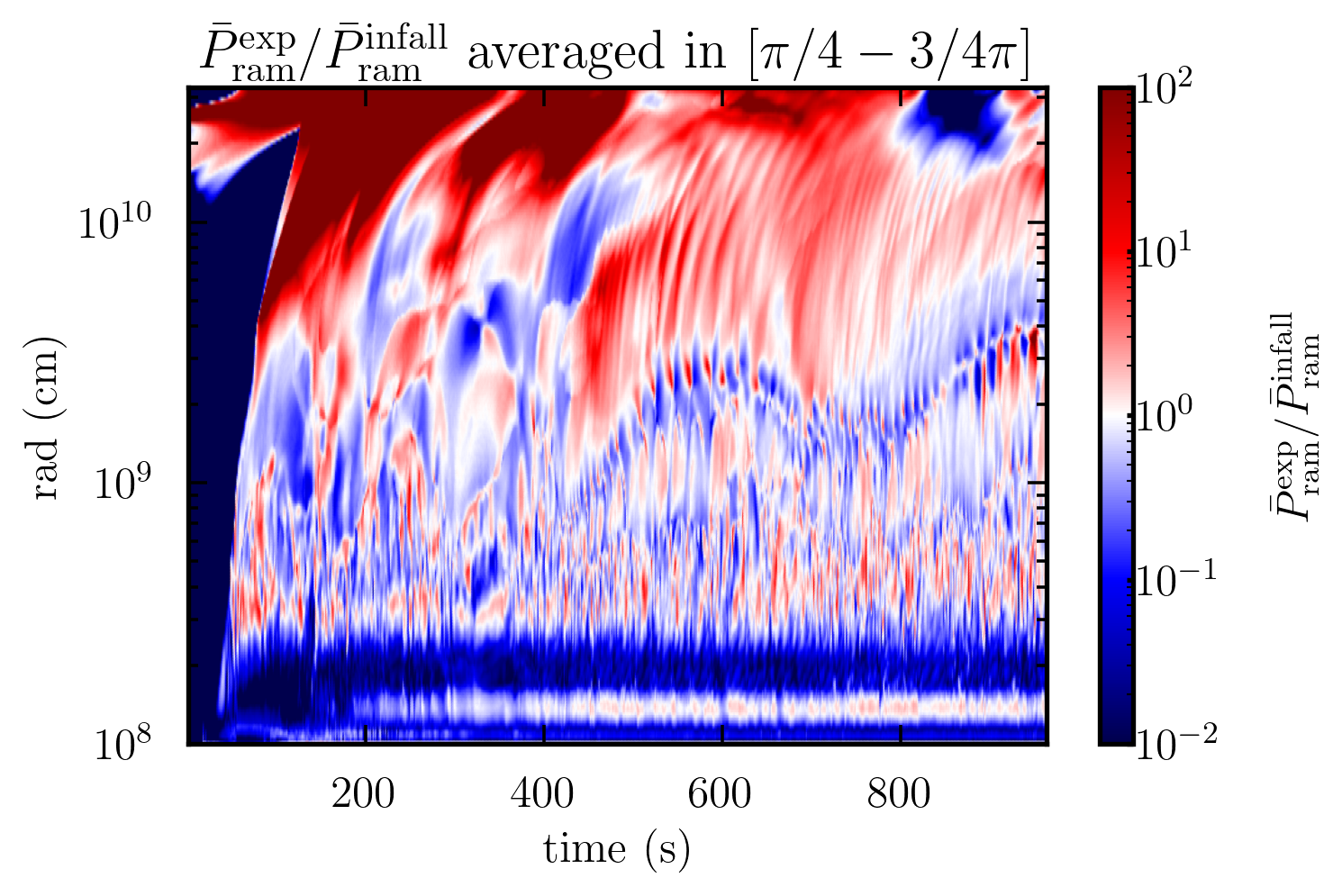}
\includegraphics [width=0.489\textwidth]{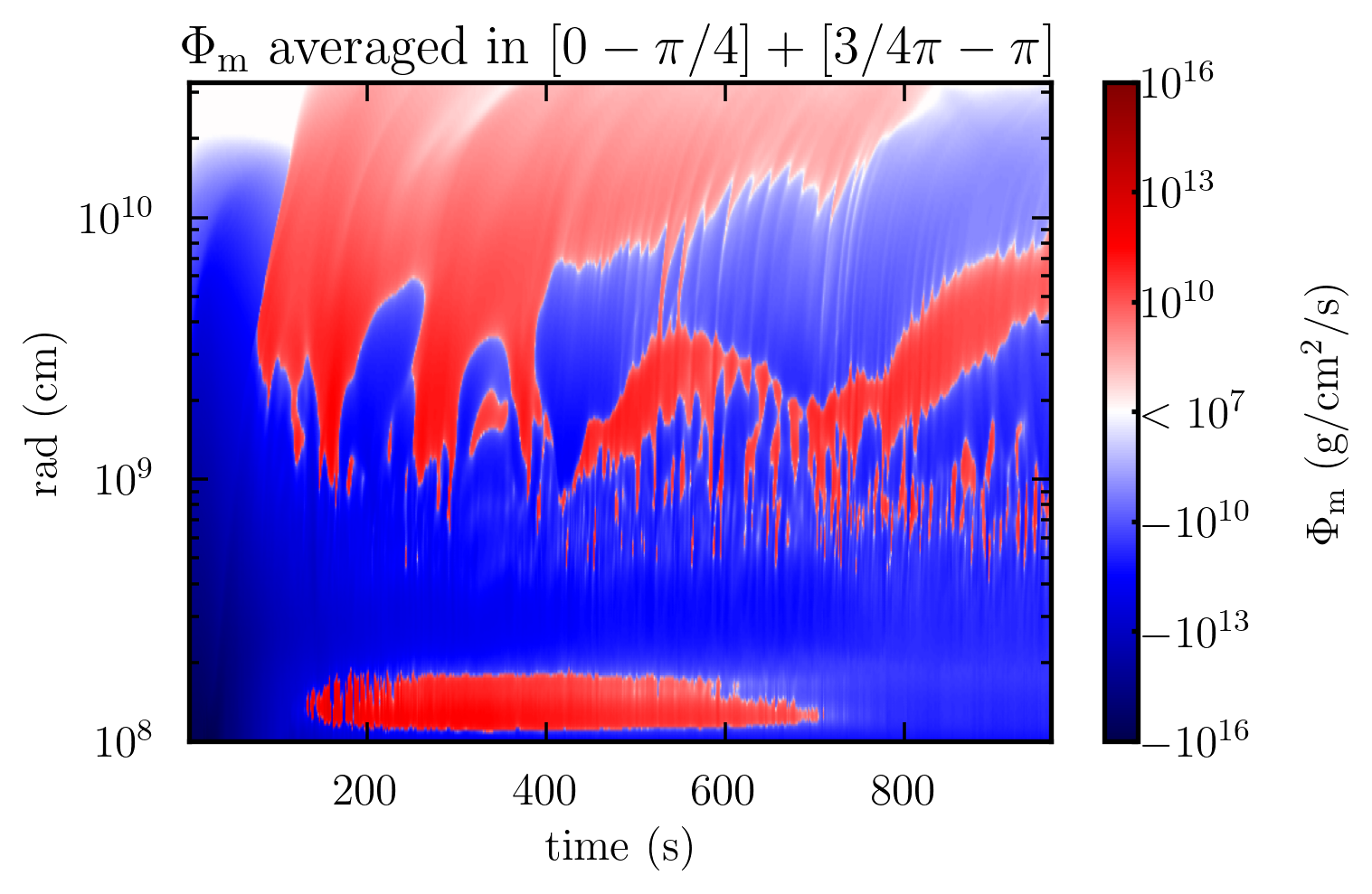}
\includegraphics [width=0.48\textwidth]{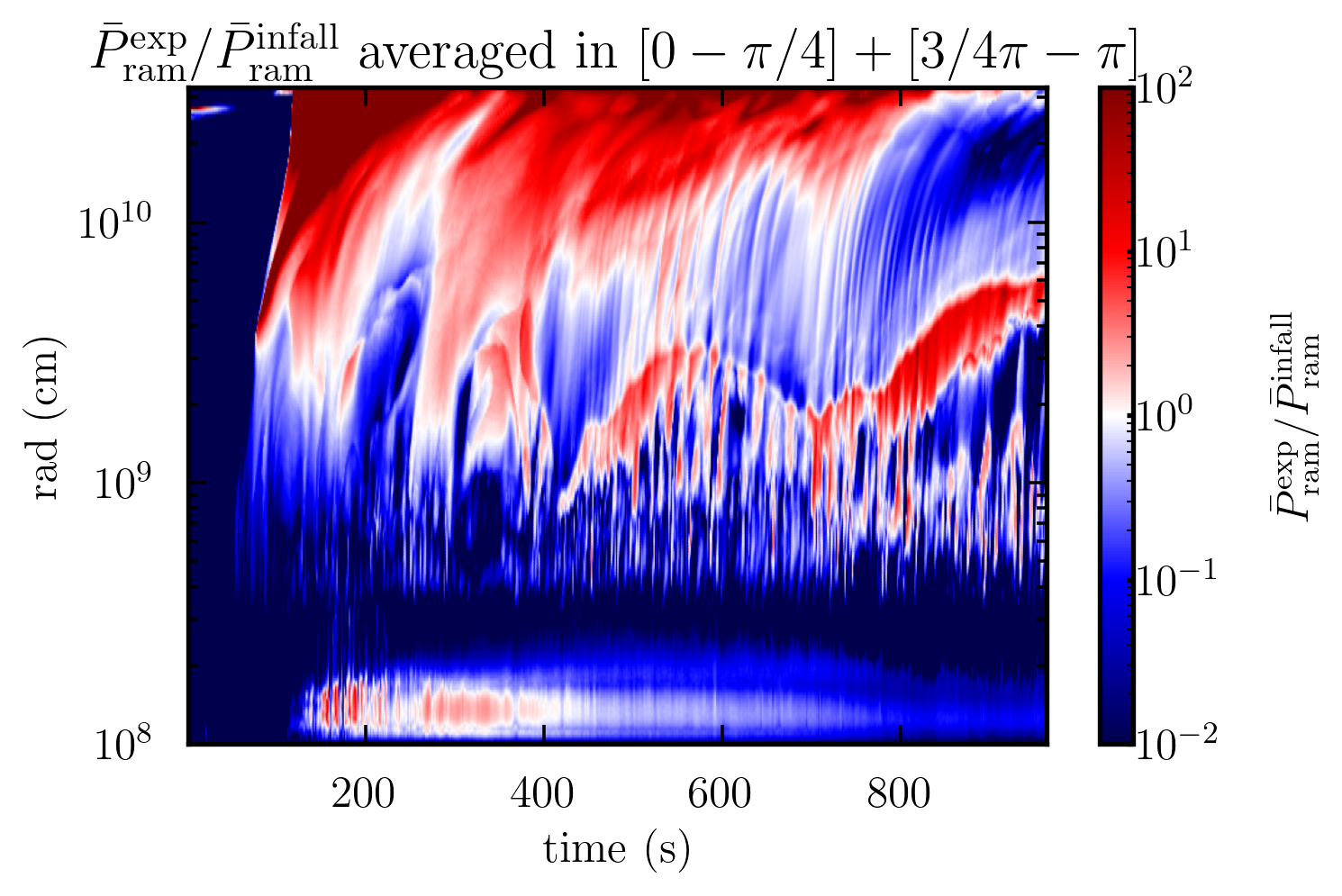}
\caption{Upper-left panel: time evolution of the mass rate $\Phi_\mathrm{m}$ averaged over the injection angle.  Upper-right panel: time evolution of the ratio between the ram pressure averaged over the injection angle of the injected matter, $\bar{P}_\mathrm{ram}^\mathrm{exp}$, and of the infalling envelope, $\bar{P}_\mathrm{ram}^\mathrm{infall}$.  Lower panels: they show the same time evolution diagrams as the upper panels but here the mass flux and the ram pressure are angled-averaged outside the injection angle, i.e., in $[0-\pi/4]+[3/4\pi-\pi]$. These plots are obtained for the model  M20\_1\_1\_0.3\_0.1 (lower panels).}
\label{fig:flux_m_Pram_spacetime_tw1}
\end{figure*}

The dynamics of the explosion of M20\_10\_1\_0.3\_0.1 can be followed in the left panel of Fig.~\ref{fig:flux_m_Pram_spacetime_tw10} where we present the time evolution of the mass outflow rate $\Phi_\mathrm{m}$ averaged over the injection angle. The angular-averaged mass outflow rate is given by:
\begin{align}
\Phi_\mathrm{m}(r)=\frac{\int_{\theta_1}^{\theta_2}\sin\theta\cdot \rho(r,\theta) v_r(r,\theta)\, d\theta }{\int_{\theta_1}^{\theta_2}\sin\theta d\theta},
\end{align}

where $\theta_1$ and $\theta_2$ are the edges of the angle within which we average the mass outflow rate, in this case they limit the injection angle between $\pi/4$ and $3/4\pi$, corresponding to $\theta_1^*$ and $\theta_2^*$ introduced in Sec.\ref{sec:Inner_Boundary_Cond}.  
We focus here on the mass outflow rate averaged over the injection angle because it results to be almost equivalent to that averaged over the entire computational region (i.e., over $[0-\pi]$). This indicates that the explosion is quasi spherical. The right panel of Fig.~\ref{fig:flux_m_Pram_spacetime_tw10} displays the time evolution of the ratio between the averaged ram pressure of the expanding matter, $\bar{P}_\mathrm{ram}^\mathrm{exp}$, and that of the infalling matter, $\bar{P}_\mathrm{ram}^\mathrm{infall}$. These are defined as:

\begin{align}
\bar{P}_\mathrm{ram}^\mathrm{exp}(r)=\frac{\int_{\theta_1, v_r>0}^{\theta_2}\sin\theta\cdot P_\mathrm{ram}(r,\theta)\, d\theta}{\int_{\theta_1, v_r>0}^{\theta_2}\sin\theta d\theta} \\
\bar{P}_\mathrm{ram}^\mathrm{infall}(r)=\frac{\int_{\theta_1, v_r<0}^{\theta_2}\sin\theta\cdot P_\mathrm{ram}(r,\theta)\, d\theta}{\int_{\theta_1, v_r<0}^{\theta_2}\sin\theta d\theta }
\end{align}

The left panel highlights the formation of a strong mass outflow at about the onset of the injection starting from the inner boundary and reaching the outer boundary. The positive mass rate dominates over the infalling envelope in the outer layers for the first $\sim 200$~s of the simulation which corresponds to the time within which the ejecta mass converges. This region of the plot perfectly corresponds to that with the highest value of $\bar{P}_\mathrm{ram}^\mathrm{exp}/\bar{P}_\mathrm{ram}^\mathrm{infall}$, in the right panel. This comparison suggests that the explosion in this model is driven by the injected wind which has a ram pressure larger than that of the infalling matter at the onset of the injection.
The competition between the ram pressure of the wind and that of the infalling envelope can be analysed more in detail in the upper panel of Fig.~\ref{fig:P_ram_2_models_comparison}, where we compare the ram pressure of the expanding (solid lines) and infalling matter (dotted lines) along the equatorial plane in this simulation at $t=19$~s, soon after the injection has started. In the figure the expanding and infalling matter close to the inner boundary represents the wind component and the infalling envelope respectively. The upper panel confirms that after the injection starts, for M20\_10\_1\_0.3\_0.1, the ram pressure of the injected matter dominates. As a result, most of the injected matter can expand outward without falling back, leading to a highly-energetic explosion with $\sim\SI{e52}{erg}$.

If the wind injection is weak, the injected matter is not able to efficiently push forward the stellar envelope determining a sub-energetic explosion with $<\SI{e51}{erg}$. The dynamics of such explosion is presented in Fig.~\ref{fig:flux_m_Pram_spacetime_tw1} using the model M20\_1\_1\_0.3\_0.1. This model has the same parameters as M20\_10\_1\_0.3\_0.1 ($t_\mathrm{acc}/t_\mathrm{w} = 1$, $\xi^2 = 0.3$ and $f_{\mathrm{therm}} = 0.1$) apart from the wind timescale which is $t_\mathrm{w} = 1$~s (and hence $t_\mathrm{acc} = 1$~s). In this simulation we measure lower injected and explosion energy, i.e. $E_\mathrm{inj}=3.08\times10^{51}$~erg and $E_\mathrm{expl}=0.088\times 10^{51}$~erg. In the upper and bottom panels of Fig.~\ref{fig:flux_m_Pram_spacetime_tw1}  we show the time evolution of $\Phi_\mathrm{m}$ (left panels)  and $\bar{P}_\mathrm{ram}^\mathrm{exp}/\bar{P}_\mathrm{ram}^\mathrm{infall}$ (right panels) averaged over the injection angle and outside that, respectively. %Unlike the highly-energetic model, in this case the explosion is not spherically symmetric, therefore in the bottom panels we also present the time evolution of these quantities averaged outside the injection angle.
The dynamics of the explosion of M20\_1\_1\_0.3\_0.1 looks different from that of M20\_10\_1\_0.3\_0.1 (see Fig.~\ref{fig:flux_m_Pram_spacetime_tw10}). In this case there is no positive $\Phi_\mathrm{m}$ dominating at all radii from the inner boundary to the outer boundary within the injection angle (see left panels of Fig.~\ref{fig:flux_m_Pram_spacetime_tw1}). This means that most of the injected matter cannot expand outwards without falling back. Indeed the upper-left panel of Fig.~\ref{fig:flux_m_Pram_spacetime_tw1} shows that a negative averaged mass rate always dominates the innermost region around $\SI{2e8}{cm}$, within the injection angle. This infalling mass stops the expansion of the injected matter. However a positive $\Phi_\mathrm{m}$ is present at larger radii. 
The bottom-left panel of Fig.~\ref{fig:flux_m_Pram_spacetime_tw1} shows that the expanding component of the mass flux seems to dominate not only at $r\gtrsim\SI{5e8}{cm}$, but also in the innermost region outside the injection angle, between $\sim180$~s and $\sim 700$~s. Nonetheless, the expanding matter is blocked by the infalling matter at around $\SI{2e8}{cm}$ here as well. These regions of dominating positive (negative) $\Phi_\mathrm{m}$ in the left panels correspond to regions in which the ram pressure of the expanding (infalling) matter dominates in the right panels. This confirms that the dominating component of the ram pressure determines the direction of the muss flux, as suggested for M20\_10\_1\_0.3\_0.1. Therefore, since in M20\_1\_1\_0.3\_0.1 the ram pressure of the injected material close to the inner boundary seems to be on average always weaker than that of the infalling, contrary to what happens in M20\_10\_1\_0.3\_0.1, the explosion in this model is unlikely to be driven by the wind.
The competition between the ram pressure of the injected wind and that of the infalling matter of this model is further investigated in the lower panel of Fig.~\ref{fig:P_ram_2_models_comparison}.
In this case even though we find some matter with positive velocity, the injected matter has a ram pressure smaller than that of the infalling envelope and some of it is found also at smaller radii, i.e. $\sim10^8$~cm, among the wind, limiting the explosion energy.

These results leads to the conclusion that if the ram pressure of the injected matter is smaller than that of the infalling matter, the explosion is determined by another mechanism: the infalling envelope bounces on the wind, launching the shock wave propagating outward. The outer layer of the star is then swept by the shock wave, being unbound. In this case, the energy source of the explosion is not the energy injected, but the released gravitational binding energy of bounced matter. Such a case is shown in the lower panel of Fig.~\ref{fig:P_ram_2_models_comparison} for the model M20\_1\_1\_0.3\_0.1. The plot shows that the ram pressure of the infalling envelope (dashed line) dominates over that of the injected matter (solid line). However the amount of the injected matter is sufficient to act as a ``wall'' causing the bounce of the infalling envelope, which occurs at around $r\lesssim\SI{e9}{cm}$. In this figure the shock wave launched is also visible at larger radii, i.e. at $r=\SI{2e9}{cm}$, propagating outwards and then sweeping the outer layer of the star.
Since this unbound mass is located only at radii of tens of thousands of kilometer, where the density is small, also the amount of the unbound mass is small.
At the same time the shock propagation decreases the infalling matter velocity and its ram pressure so that if the latter becomes sufficiently small, then the injected matter can move over it and go outwards becoming another ejecta components. However this happens after hundreds of seconds, when the energy budget - which is determined by the mass supply of the infalling envelope to the disk - is low determining also a very low $E_\mathrm{inj}$.

The comparison of the $P_\mathrm{ram}$ competition between the two models shows that whether the ram pressure of the disk wind can overcome the ram pressure of the accretion flow or not determines a distinction between highly-energetic explosions and sub-energetic explosions. The efficiency of the explosion mechanism can also be measured by the ratio $E_\mathrm{expl}/E_\mathrm{inj}$ which indicates the fraction of the injected energy transferred to the ejecta. For M20\_10\_1\_0.3\_0.1, the explosion energy is the $\sim 64\%$ of the injected energy, while in the case of M20\_1\_1\_0.3\_0.1, $E_\mathrm{expl}/E_\mathrm{inj}\approx 0.029$.

We find that some models with sub-energetic explosions have $E_\mathrm{expl} > E_\mathrm{inj}$. For example, model M20\_1\_1\_0.1\_0.01 has $E_\mathrm{expl}\approx\SI{5e49}{erg}$, while $E_\mathrm{inj}<\SI{e49}{erg}$ (see Table.~\ref{tab:models}). 
This can happen because the energy source of such sub-energetic explosion is different from that of the injected matter, it is the gravitational binding energy released by the bouncing infalling envelope (see Sec.~\ref{sec:hydrodynamics}).

\subsection{Bimodality of $E_\mathrm{expl}$}\label{sec:model_comparison} 

\begin{figure*}
\centering
	%\captionsetup{font=small, labelfont=rm}
	%\subfloat[]
	\includegraphics [width=0.48\textwidth]{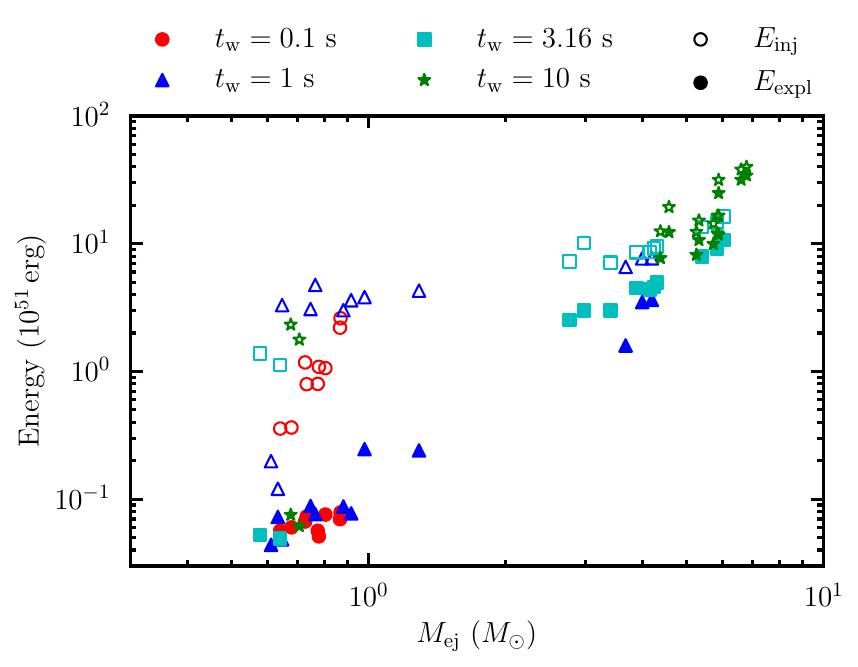}	
		%\label{fig:Mej_Eexpl} 
	\includegraphics [width=0.48\textwidth]{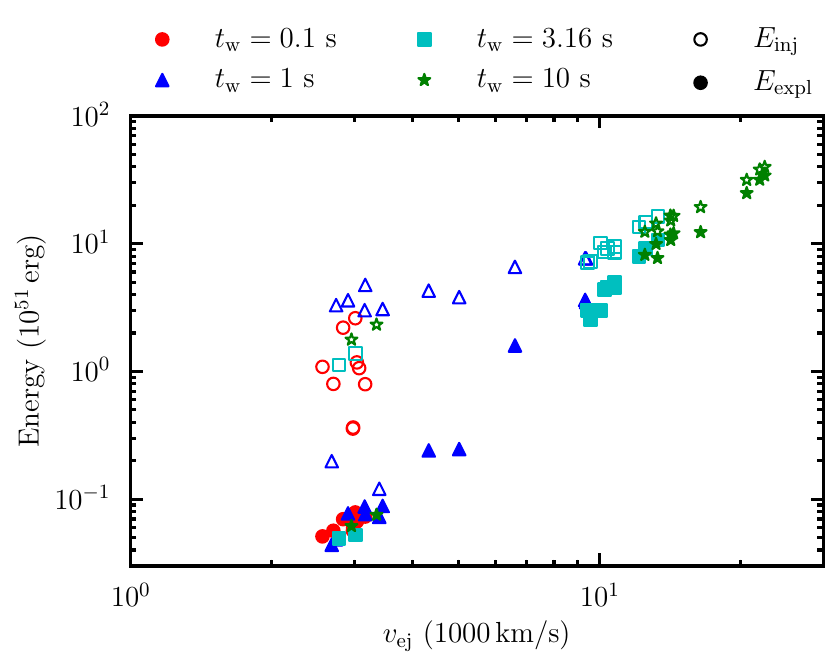}
		%\label{fig:vej_Eexpl} 
\caption{The $E_\mathrm{expl}$ (filled markers) and the $E_\mathrm{inj}$ (open markers) against the  ejecta mass $M_\mathrm{ej}$ (left panel) and against the ejecta velocity $v_\mathrm{ej}$ (right panel) for the models studied in this work. Results for models with different wind time scales $t_\mathrm{w}=$ 0.1s, 1s, 3.16s and 10s are distinguished by different colours, respectively red, blue, cyan and green.
	}
\label{fig:Mej_vej_Eexpl}
%\label{fig:P_ram_charact_model}
\end{figure*}

We plot the injected and explosion energies of all models with different parameters against respectively the ejecta mass (left panel of Fig.~\ref{fig:Mej_vej_Eexpl}) and the average ejecta velocity  $v_\mathrm{ej}=\sqrt{2E_\mathrm{expl}/M_\mathrm{ej}}$ (right panel of Fig.~ \ref{fig:Mej_vej_Eexpl}).
The figures show that the models fall into two categories having different explosion energies, while such a distinction is not seen for the injection energy, which shows a rather continuous distribution.
The first category is made of highly-energetic explosions, characterized by the explosion energy of about $10^{52}$ erg. The second category is made of sub-energetic explosions with an energy of approximately $10^{50}$ erg. 
This bimodal distribution seems to be mainly controlled by the wind time scale. 
Models with shorter  $t_\mathrm{w}$ (i.e. $t_\mathrm{w}\sim 0.1-1$~s) are located in the left, low-energy side of the plots, while those with longer $t_\mathrm{w}$ ($\sim 3.16-10$~s) belong to the high explosion energy group, on the right of our plots. 

Another feature distinguishing the highly-energetic and sub-energetic explosions is the fraction of the injected energy converted in explosion energy which can be measured by the ratio $E_\mathrm{expl}/E_\mathrm{inj}$.  Both panels in Fig.~\ref{fig:Mej_vej_Eexpl} shows that the gap between $E_\mathrm{inj}$ and $E_\mathrm{expl}$ decreases with increasing $M_\mathrm{ej}$ and $v_\mathrm{ej}$, indicating a correlation between $E_\mathrm{expl}/E_\mathrm{inj}$  and $E_\mathrm{expl}$.

\begin{figure}
	\centering
	\captionsetup{font=small, labelfont=it}
	%\subfloat[]
	{\includegraphics [width=0.48\textwidth]{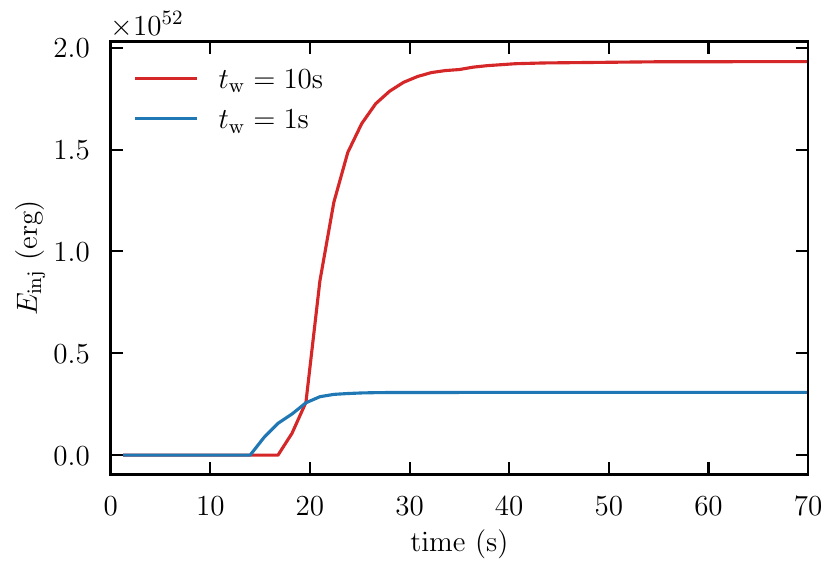}	
		\label{fig:E_injected_comparison} }\quad
	%\subfloat[]
	{\includegraphics [width=0.48\textwidth]{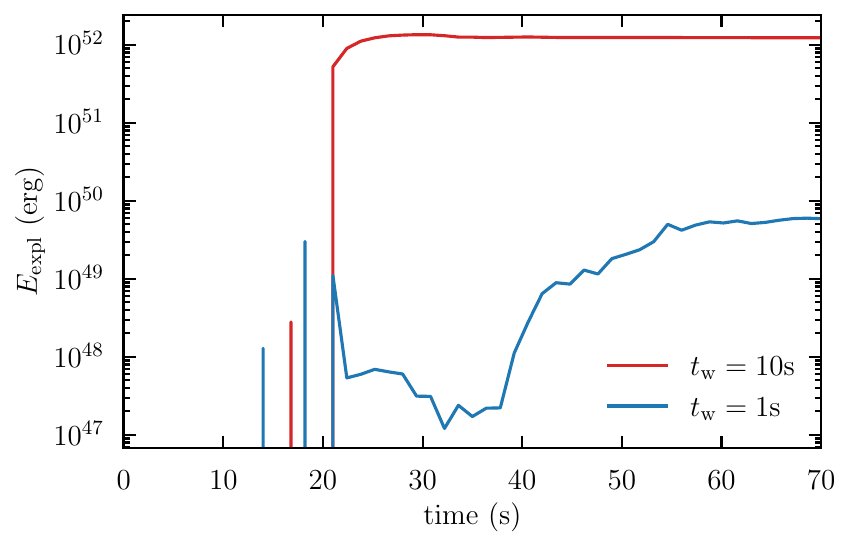}
		\label{fig:E_expl_evol_comparison} }
	\caption{Upper panel: evolution of the injected energy in the first 70 seconds of the simulation for the models M20\_10\_1\_0.3\_0.1 and M20\_1\_1\_0.3\_0.1. Both models have the same parameters apart from $t_\mathrm{w}$ (see Table~\ref{tab:models}).  Model M20\_10\_1\_0.3\_0.1 is represented using red lines, while model M20\_1\_1\_0.3\_0.1 is represented using blue lines. Lower panel: evolution of the explosion energy in the first 70 seconds of the simulation for the same two models.}
\label{fig:E_injected_expl_comparison}
\end{figure}

To explain the bimodality of the explosion, we compare the evolution of two models belonging to the highly-energetic and sub-energetic categories respectively. Since M20\_10\_1\_0.3\_0.1, described in Section~\ref{sec:hydrodynamics}, lays in the right-hand side of both panels of Fig.~\ref{fig:Mej_vej_Eexpl}, we compare it to the model M20\_1\_1\_0.3\_0.1 having the same parameters ($t_\mathrm{acc}/t_\mathrm{w} = 1$, $\xi^2 = 0.3$ and $f_{\mathrm{therm}} = 0.1$), but  a different $t_\mathrm{w} = 1$~s (and hence $t_\mathrm{acc} = 1$~s) which is instead located on the left, low-energy side.
In Fig.~\ref{fig:E_injected_expl_comparison}, we compare the time evolution of injected energy (upper panel) and explosion energy (lower panel) of the two models in the first 70 seconds of the simulation. 
Compared to the injected and explosion energies of M20\_10\_1\_0.3\_0.1 (see Sec.~\ref{sec:hydrodynamics}), M20\_1\_1\_0.3\_0.1 reaches only $E_\mathrm{inj}=3.08\times 10^{51}$ erg and $E_\mathrm{expl} = 0.088\times 10^{51}$ erg (these values are also shown in Table~\ref{tab:models}).
In the model with $t_\mathrm{w}=10$ s the injected energy grows faster with a steeper slope after the onset of the injection.

\begin{figure}
	\centering
	\captionsetup{font=small, labelfont=it}
	%\subfloat[]
	{\includegraphics [width=0.48\textwidth]{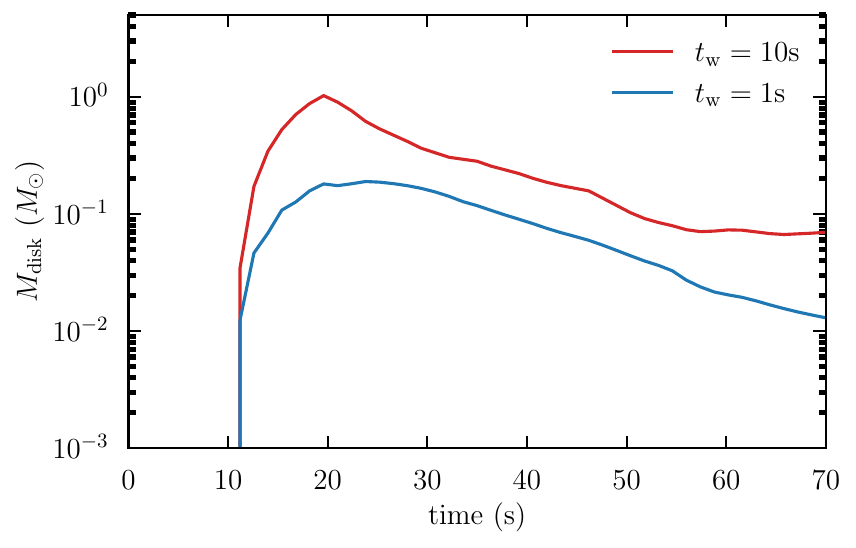}	
		\label{fig:M_disk_comparison} }\quad
%	\subfloat[]
	{\includegraphics [width=0.48\textwidth]{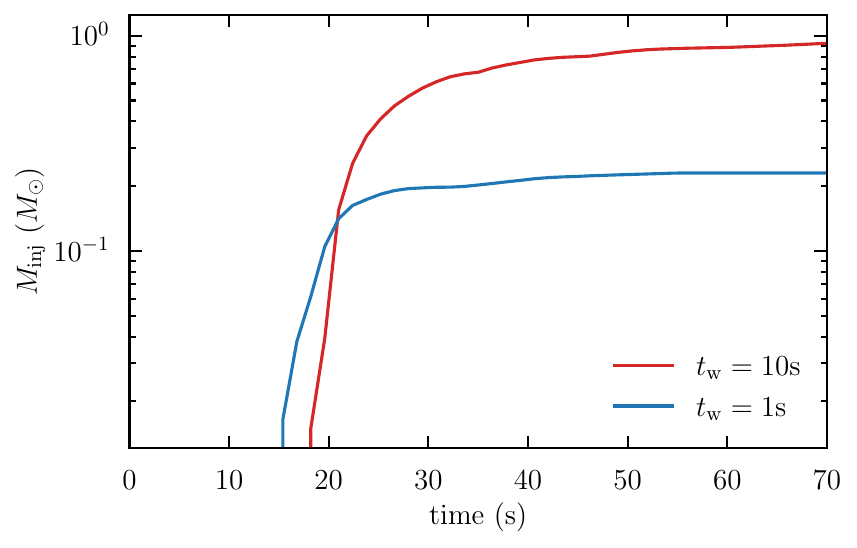}
		\label{fig:M_inj_comparison} }
	\caption{Upper panel: evolution of the disk mass in the first 70 seconds for the simulation of the same models as in Fig.~\ref{fig:E_injected_expl_comparison}. Lower panel: evolution of the injected mass in the first 70 seconds of the simulation for the same two models. }
\label{fig:M_disk_M_inj_comparison}
\end{figure}

\begin{figure}
	\centering
	\includegraphics [width=0.48\textwidth]{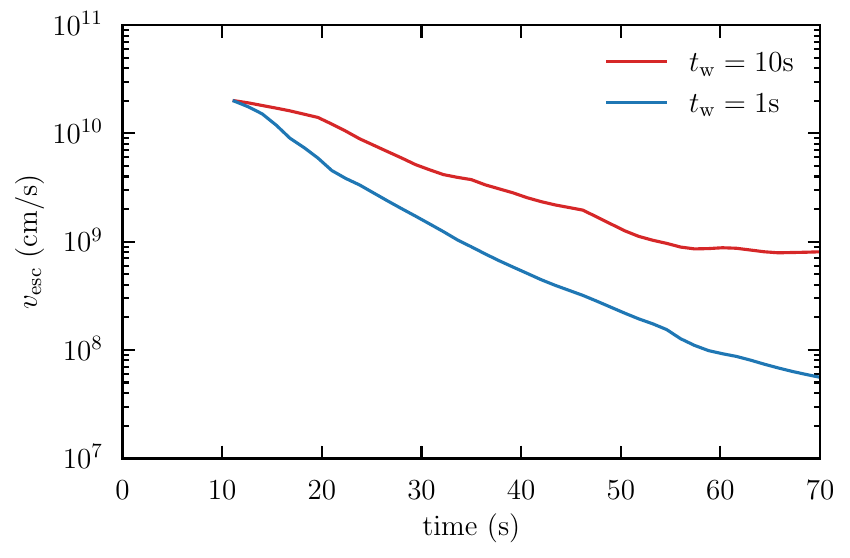}
	\caption{Comparison of the escape velocity between the same models as in Fig.~\ref{fig:E_injected_expl_comparison}.}
    \label{fig:v_esc_comparison_with_charact_model}
\end{figure}

It is clear from equation~\eqref{M_dot_wind} that, for a given disk mass, the shorter wind time scale leads to the higher mass injection rate. The upper panel of Fig.~\ref{fig:M_disk_M_inj_comparison} shows that the difference in the disk mass is always less than an order of magnitude in the first $\SI{70}{s}$ of the evolution. Since the wind time scales are different by a factor of ten for those two models, the mass injection rate in model M20\_1\_1\_0.3\_0.1 may be always larger than that in M20\_10\_1\_0.3\_0.1. Indeed, the lower panel of Fig.~\ref{fig:M_disk_M_inj_comparison} shows that the mass injection begins slightly earlier in model M20\_1\_1\_0.3\_0.1 than the other model, which indicates that the wind power is stronger in the model in the earlier phase. 
However, the mass injection rate is somehow smaller in the later phase ($t\gtrsim\SI{20}{s}$). 
This stems from the smaller escape velocity of the disk, $v_\mathrm{esc}$, as found in Fig.~\ref{fig:v_esc_comparison_with_charact_model}, especially in the later phase. The escape velocity plays an important role in determining the mass injection rate and affects it more than $t_\mathrm{w}$ because the  flux $\dot{M}_\mathrm{wind}$ defined in equation~\eqref{M_dot_wind} is actually computed according to equation~\eqref{rho_from_M_wind} solving the Riemann problem at the inner boundary.\footnote{Solving the Riemann problem at the inner boundary may be an additional reason for $\dot{M}_\mathrm{disk}<M_\mathrm{disk}/t_\mathrm{w}$. In addition, the discrepancy of the computed $\dot{M}_\mathrm{disk}$ from $M_\mathrm{disk}/t_\mathrm{w}$ is larger for smaller escape velocity}
The smaller escape velocity leads to the smaller specific energy of injected matter, which is proportional to $\xi^2 v_\mathrm{esc}^2$.
Hence, the wind is injected less efficiently in model M20\_1\_1\_0.3\_0.1 in the later phase. Therefore, the injected mass and energy saturate earlier.

As found in equation~\eqref{v_esc}, which can be rewritten as $v_\mathrm{esc} =\sqrt{2} GM_\mathrm{BH}/j_\mathrm{disk}$ with equation~\eqref{r_disk_definition}, the larger specific angular momentum of the disk leads to smaller escape velocity. 
Using Eqs.~\eqref{M_dot_disk}, \eqref{J_dot_disk}, and \eqref{M_BH_accretion}--\eqref{J_dot_wind}, the evolution equation of the disk specific angular momentum is written as:
%\begin{align}\label{dJ_disk_over_dM_disk}
 %   \frac{d \ln(j_\mathrm{disk})}{dt} =\frac{1}{t_\mathrm{acc}}\bigg( 1-\frac{j_\mathrm{ISCO}}{j_\mathrm{disk}} \bigg)  + \bigg(\frac{j_\mathrm{fall}}{j_\mathrm{disk}} - 1\bigg)\frac{\dot{M}_\mathrm{fall,disk}}{M_\mathrm{disk}},
%\end{align}

\begin{align}\label{dJ_disk_over_dM_disk}
    \frac{d j_\mathrm{disk} }{dt}
    =   \frac{1}{ t_\mathrm{acc} }( j_\mathrm{disk}-j_\mathrm{ISCO} )  
    +   \frac{\dot{M}_\mathrm{fall,disk}}{M_\mathrm{disk}} (j_\mathrm{fall}-j_\mathrm{disk}) ,
\end{align}
where $j_\mathrm{fall}:=\dot{J}_\mathrm{fall,disk}/\dot{M}_\mathrm{fall,disk}$ is the specific angular momentum of the infalling matter. The two terms of the equation are the contribution of the mass accretion onto the BH and the contribution of the infalling envelope respectively. 
Due to the contribution of the first term, if it dominates over the second, the disk specific angular momentum $j_\mathrm{disk} = J_\mathrm{disk}/M_\mathrm{disk}$ always increases because $j_\mathrm{disk}>j_\mathrm{ISCO}$. 
The second term does not always add a positive contribution to $j_\mathrm{disk}$ since it can be also negative when $j_\mathrm{disk}>j_\mathrm{fall}$. If the absolute value of these negative contributions is smaller that the first term, then the accretion onto the BH dominates and $j_\mathrm{disk}$ keeps increasing.  In our simulations we find that the second term becomes also negative, i.e. $j_\mathrm{disk}>j_\mathrm{fall}$, however its absolute value is on average smaller that the first term. In this case, it is evident from the equation that the time scale of the increase in $j_\mathrm{disk}$ is determined by $t_\mathrm{acc}$. Therefore, the decrease in $v_\mathrm{esc}$ and thus the energy injection efficiency drop faster in model M20\_1\_1\_0.3\_0.1 than those in model M20\_10\_1\_0.3\_0.1.

\subsection{Parameter dependence of $E_\mathrm{expl}$ and $M_\mathrm{ej}$}\label{sec:parameter_dependence}
\begin{figure}
	\centering
	\includegraphics [width=0.48\textwidth]{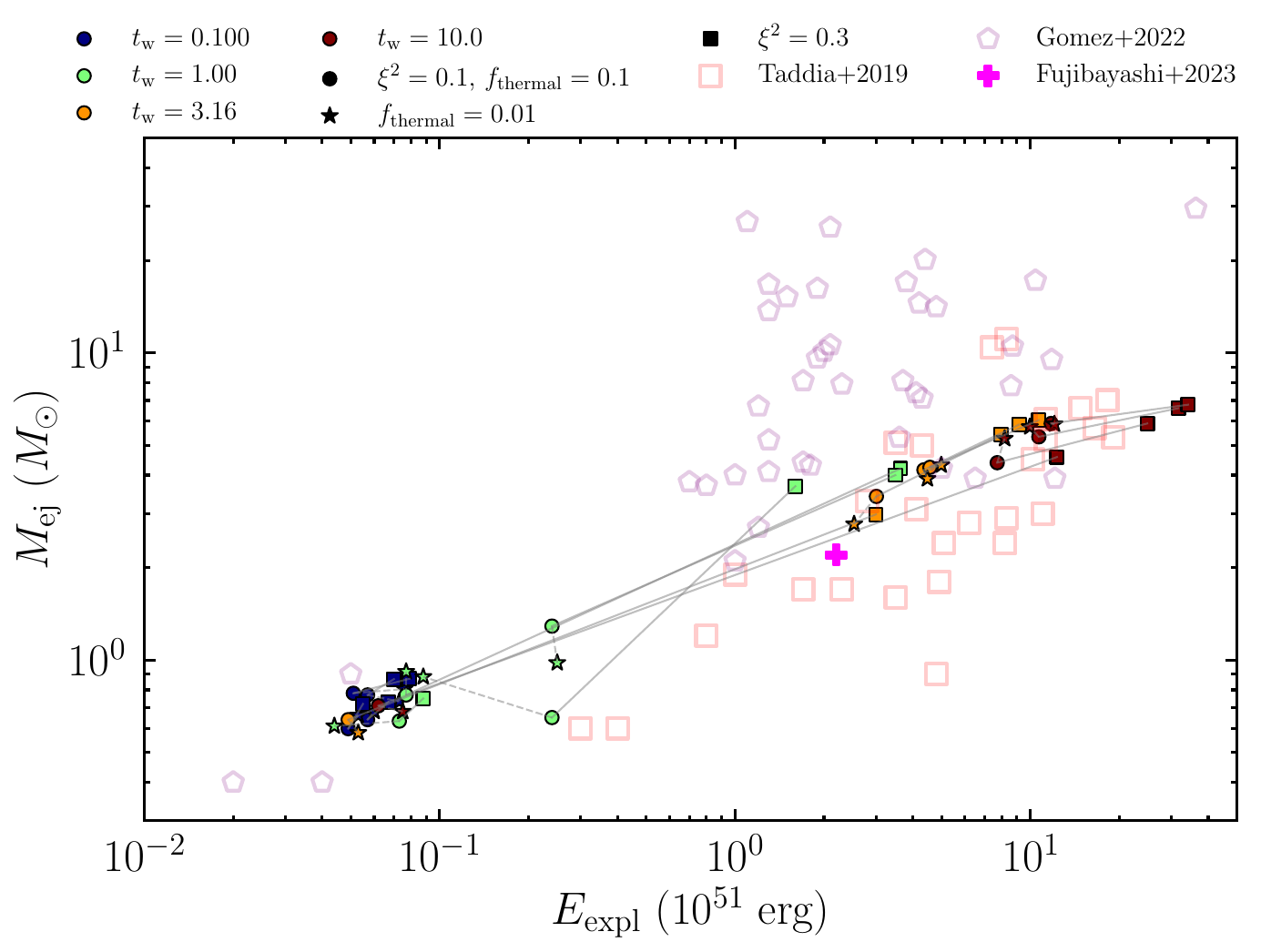}
	\caption{Parameter dependence with respect to the observable pair of ejecta mass $M_\mathrm{ej}$ and explosion energy $E_\mathrm{expl}$. The color distinguishes the wind time scale $t_\mathrm{w}$. The lines connect models with  all the other parameters fixed but different $\xi^2$ or $f_\mathrm{therm}$. The open markers display the observational data for stripped-envelope SNe, some of which are broad-lined type Ic SNe, taken from \citet{Taddia2019jan} and \citet{Gomez2022dec}. The magenta plus-sign denotes the result obtained in a general relativistic neutrino-radiation viscous-hydrodynamics simulation with the same progenitor \citep{Fujibayashi2023arxiv}.}
    \label{fig:Eexp_vs_Mej_tw_parameter_dependence}
\end{figure}

In this section we analyse the effects that the parameters of our model  have on the explosion. To do that we show in Fig.~\ref{fig:Eexp_vs_Mej_tw_parameter_dependence} the distribution of our models in the $E_\mathrm{expl}$-$M_\mathrm{ej}$ plane. We display our results (filled markers) together with the  observational data for broad-lined type Ic SNe taken from \citet{Taddia2019jan} and  for stripped-envelope SNe and superluminous SNe from \citet{Gomez2022dec} (empty markers). 
We also show the results obtained by \citet{Fujibayashi2023arxiv} who did a first-principled general relativistic neutrino-radiation viscous-hydrodynamics simulation using the same progenitor model.
The values they measured for the ejecta mass of $M_\mathrm{ej}= 2.2\, M_\odot$ and the explosion energy of $E_\mathrm{expl}=2.2\times10^{51}$ erg can be considered as the lower limits for the self-consistent simulation since they are still growing at the end of their simulation.
The wind time scale of our models is indicated by the color of the marker, while the shape distinguishes models with different $\xi^2$ and $f_\mathrm{therm}$. The lines connect models with the same $t_\mathrm{w}$ and $t_\mathrm{acc}$, but different $\xi^2$ or $f_\mathrm{therm}$.

The first parameter we study is the wind time scale,  $t_\mathrm{w}$, sampled in the interval $(0.1, 1, 3.16, 10)$ s.
Fig.~\ref{fig:Eexp_vs_Mej_tw_parameter_dependence} shows that most of the models with $t_\mathrm{w}\ge 3.16$~s lay on the right side of the panel having $E_\mathrm{expl}\gtrsim 1\times10^{51}$~erg and $M_\mathrm{ej}\gtrsim2.5\,M_\odot$, while all models with $t_\mathrm{w}=0.1$~s  occupy the lower-left side of the plot presenting $E_\mathrm{expl}\lesssim  10^{51}$~erg and $M_\mathrm{ej}\lesssim 0.07\,M_\odot$. Longer wind time scales lead to higher explosion energy because they keep a larger escape velocity for longer time, as explained above and shown in Fig.~\ref{fig:v_esc_comparison_with_charact_model}. Only in models with $t_\mathrm{w}= 1$~s this parameter seems not to be predominant with respect to the $\xi^2$ and $f_\mathrm{therm}$ in determining whether an explosion in highly or sub-energetic. This case and the difference with models having other $t_\mathrm{w}$ will be discussed in a following paragraph, after analysing the general effects of the two other parameters $\xi$ and $f_\mathrm{therm}$.

Considering $\xi^2$, the figure shows that a higher value of the parameter increases the explosion energy and the mass ejecta. This effect is evident by following the gray solid lines in Fig.~\ref{fig:Eexp_vs_Mej_tw_parameter_dependence} from the circular to the squared markers which represent models with all the same parameter but $\xi^2=0.1$ and $\xi^2=0.3$ respectively. The lines show that in all simulation an increase in $\xi^2$ makes the point move towards the upper-right side of the plot, i.e. towards higher ejecta mass and explosion energy. This is consistent with our model of the wind (see equation~\eqref{e_inj_eqs}) in which we use $\xi^2$ to set the asymptotic velocity of the injected matter as $\xi v_\mathrm{esc}$. 
Hence, a higher $\xi$ corresponds to a larger kinetic energy of the wind and enhances the energy budget for the explosion.
Similarly, $E_\mathrm{expl}$ and $M_\mathrm{ej}$ increase by decreasing $f_\mathrm{therm}$.
According to equation~\eqref{e_inj_eqs}, $f_{\mathrm{therm}}$ determines the ratio between the internal to the kinetic energies of the wind. Since the sum of these energies is provided by $ (1/2) \xi^2 v_{\mathrm{esc}}^2$, reducing $f_{\mathrm{therm}}$ increases the kinetic energy by decreasing the fraction of the thermal energy.

Observing the distribution of our numerical results, we note that, despite the variation of the parameters in wide ranges, the explosion of the $M_\mathrm{ZAMS}=20\,M_\odot$ tend to remain located along a line and it does not spread to cover the space in the $M_\mathrm{ej}-E_\mathrm{expl}$ as the observational data do. In other words, the correlation between the explosion energy and the ejecta mass is tighter in our simulations than that observationally measured by \citet{Taddia2019jan}  and \citet{Gomez2022dec}.

We also note that the results from our simulations presented here are in good agreement with that obtained in \citet{Fujibayashi2023arxiv}, especially considering that their values of $E_\mathrm{expl}$ and $M_\mathrm{ej}$ are still growing at the end of their simulation, so higher values were expected if they ran the simulation longer, which will be comparable to our results. This indicates that our model contains the case studied in \citet{Fujibayashi2023arxiv} and explores various possibilities with a variety of wind properties.

\begin{figure*}
\centering
\includegraphics [width=0.48\textwidth]{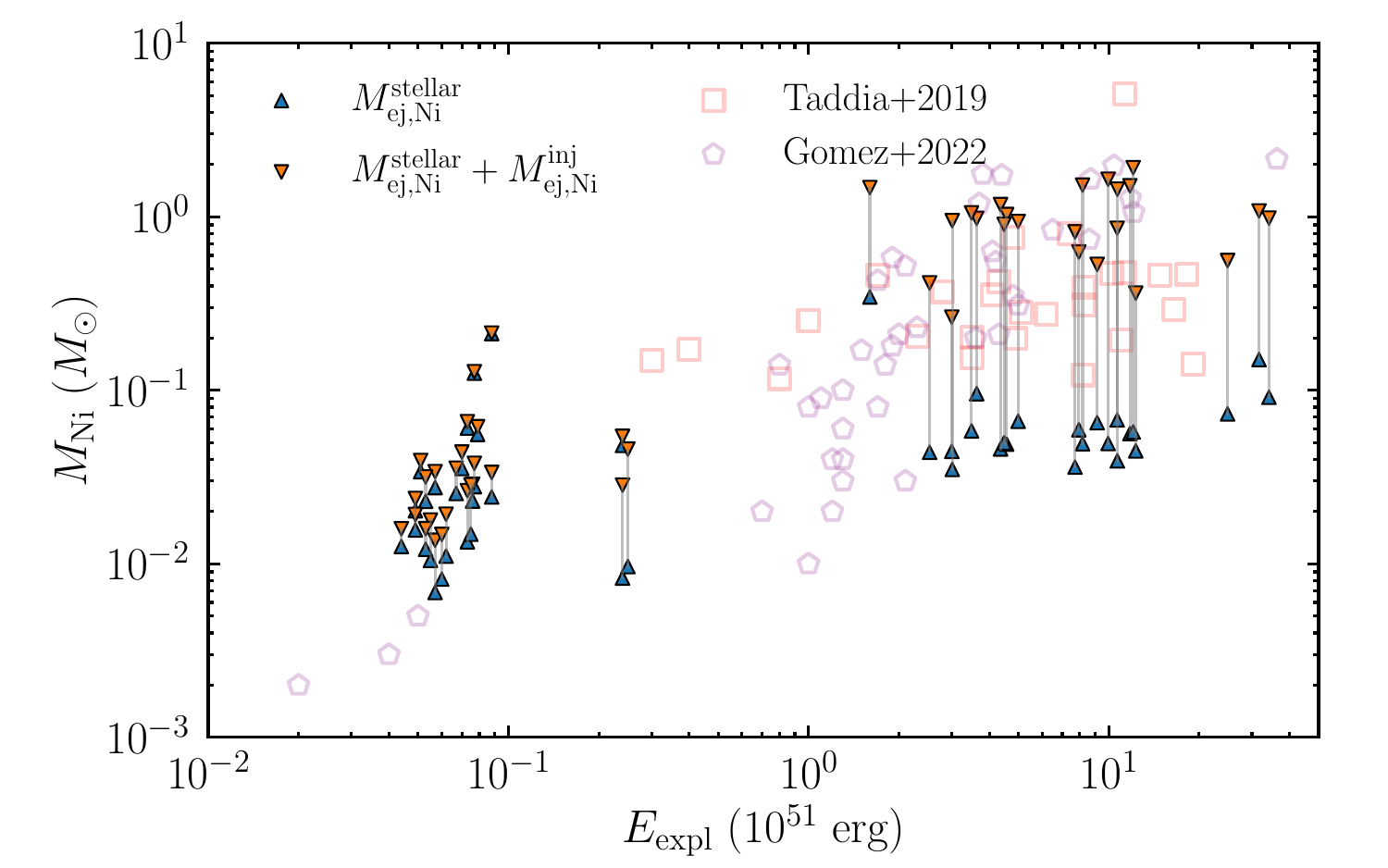}
\includegraphics [width=0.48\textwidth]{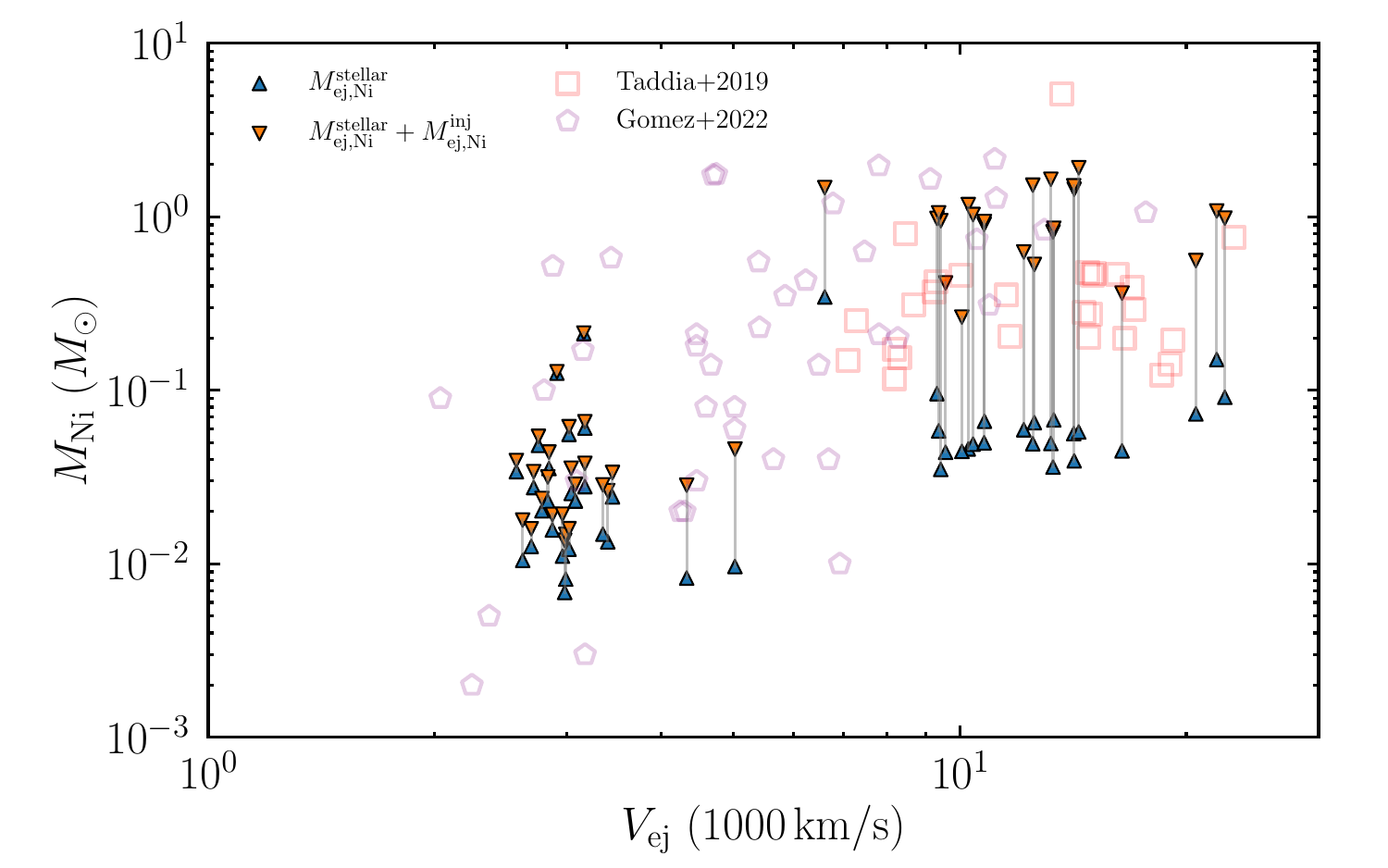}
\caption{Relations between the explosion energy and the $^{56}$Ni mass (left) and average velocity of the ejecta and the $^{56}$Ni mass (right). Each grey line connects $M^\mathrm{stellar}_\mathrm{ej,Ni}$ (triangles) and $M^\mathrm{stellar}_\mathrm{ej,Ni} + M^\mathrm{inj}_\mathrm{ej}$ (down-pointing triangles) of the same model to show the possible range of $^{56}$Ni mass. The yellow stars denote those obtained in a general relativistic neutrino-radiation viscous-hydrodynamics simulation with the same progenitor \citep{Fujibayashi2023arxiv}. The open markers display the observational data for stripped-envelope SNe, some of which are broad-lined type Ic SNe, taken from \citet{Taddia2019jan} and \citet{Gomez2022dec}.}
\label{fig:Ni_observational_data}
\end{figure*}

\subsection{$^{56}$Ni production}\label{sec:Ni_production}

%In the second part of this work we study the production of $^{56}$Ni and its dependency on the parameters of our models to check if we can produce the amount of $^{56}$Ni measured in hypernovae.

In this section, we then present the results of the $^{56}$Ni production in our models and its dependence on the free parameters of the simulations to understand if it is possible to reproduce the observational data like those presented by \citet{Taddia2019jan} and \citet{Gomez2022dec}, especially for the high-energy SN ($E_\mathrm{expl}>10^{52}$ erg). 

The results of our nucleosynthesis calculations are presented in Table~\ref{tab:models}, where we list: the $^{56}$Ni mass synthesized in the stellar component of the ejecta (the part of the ejecta originated from the computational domain),  $M^\mathrm{stellar}_\mathrm{ej,Ni}$ and the mass of ejecta component originated from the injected matter (i.e. from the disk), $M^\mathrm{inj}_\mathrm{ej}$. 
In Table~\ref{tab:models} we also show the mass of the stellar component of the ejecta reaching temperature higher than 5~GK($=\SI{5e9}{K}$), $M^\mathrm{stellar}_\mathrm{ej,>5GK}$, as the $^{56}$Ni production primarily occurs for $T\ge5$GK.

Although the injected matter explains $0.9-40$\% of the total ejecta mass, it is hard to accurately estimate the $^{56}$Ni production for this component since the complete thermodynamical history is not available.
We, therefore, estimate the upper limit of the mass of $^{56}$Ni in the ejecta, $M_\mathrm{ej,Ni}$, considering that the injected matter entirely becomes $^{56}$Ni:  $M_\mathrm{ej,Ni} = M^\mathrm{stellar}_\mathrm{ej,Ni}+M^\mathrm{inj}_\mathrm{ej}$. It is found that it ranges from $\sim0.02$ to $\sim2.09$ $M_\odot$, which corresponds to $\sim 2.2-47 \%$ of the total ejecta mass. %The $^{56}$Ni production efficiency strongly depends on the thermal history of the matter during the explosion, therefore it does not present a tight correlation with $M_\mathrm{ej}$. 

Fig.~\ref{fig:Ni_observational_data} shows $M_\mathrm{ej,Ni}$ and $M^\mathrm{stellar}_\mathrm{ej,Ni}$, the $^{56}$Ni ejecta mass that originates from the stellar component, as a function of the explosion energy (left panel) and average ejecta velocity (right panel). Together with the results of our simulations, represented by the up and down-pointing filled triangles, we also display the observational data for broad-lined type Ic SNe taken from \citet{Taddia2019jan} and for stripped-envelope SNe and superluminous SNe taken from \citet{Gomez2022dec}. We also show the $^{56}$Ni mass obtained by \citet{Fujibayashi2023arxiv}.% for the same progenitor star in a general relativistic neutrino-radiation viscous-hydrodynamics simulation, where the authors measured an ejecta mass of $M_\mathrm{ej}= 2.2\, M_\odot$ and an explosion energy of $E_\mathrm{expl}=2.2\times10^{51}$ erg.These values are lower limit for the two quantities since they are  still growing at the termination of their simulation. 

The simulation results highlight that, for most models, $M^\mathrm{inj}_\mathrm{ej}$ is likely to dominate the total $^{56}$Ni mass produced. They also show that the difference between $M_\mathrm{ej,Ni}$ and $M^\mathrm{stellar}_\mathrm{ej,Ni}$ is larger for higher explosion energies and, equivalently, for higher averaged ejecta velocities.
This means that the value $M_\mathrm{ej,Ni}$ we estimated is affected by the uncertainty of the thermodynamical history of $M^\mathrm{inj}_\mathrm{ej}$, and this is particularly true for the highly-energetic models.

Looking at Fig.~\ref{fig:Ni_observational_data} it is  found that our numerical results for the $^{56}$Ni mass  reproduce the relation between $M_\mathrm{Ni}$ and $E_\mathrm{expl}$ or $M_\mathrm{Ni}$ and $v_\mathrm{ej}$
for very-high energy SNe with $E_\mathrm{expl}>2\times 10^{51}$ erg and sub-energetic SNe with $E_\mathrm{expl}< 0.1\times10^{51}$ erg of the observational data, suggesting that these SNe may be driven by a wind-driven explosion modeled as in this work. Moreover, Fig.~\ref{fig:Ni_observational_data} shows that also considering the $^{56}$Ni mass produced, the models fall into two categories which correspond to those of highly-energetic and sub-energetic explosions observed in Fig.~\ref{fig:Mej_vej_Eexpl}.
Sub-energetic explosions produce a smaller amount of  $^{56}$Ni ($<0.2\,M_\odot$) while highly-energetic explosions produce $0.2-2.1 \,M_\odot$ of $^{56}$Ni. 

Our estimate of the $^{56}$Ni ejecta mass is in good agreement also with that obtained in \citet{Fujibayashi2023arxiv}, as it was for $E_\mathrm{expl}$ and $M_\mathrm{ej}$  (see Fig.~\ref{fig:Eexp_vs_Mej_tw_parameter_dependence}), especially considering that their values of $E_\mathrm{expl}$ and $M_\mathrm{Ni}$ were still growing at the end of their simulation. 

\begin{figure}
	\centering
	\includegraphics [width=0.48\textwidth]{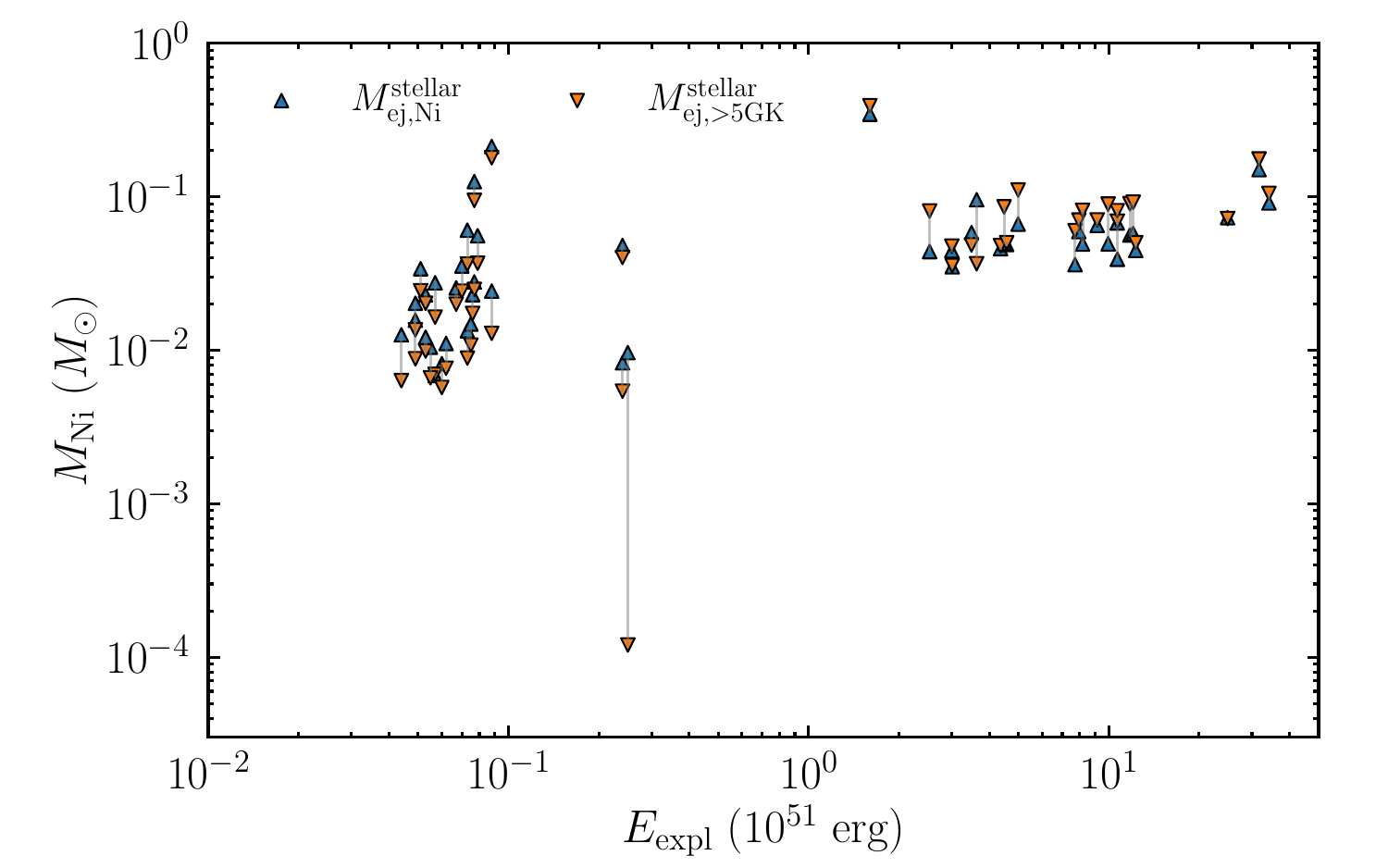}
	\caption{Comparison between the $M^\mathrm{stellar}_\mathrm{ej,Ni}$ (blue triangles) and $M^\mathrm{stellar}_\mathrm{ej,>5GK}$ (orange down-pointing triangles) displayed against the explosion energy. Each grey line connects $M^\mathrm{stellar}_\mathrm{ej,Ni}$ and $M^\mathrm{stellar}_\mathrm{ej,>5GK}$ of the same model. }
    \label{fig:M5GK_vs_MNi_only_our_data}
\end{figure}

$M_\mathrm{ej,>5GK}$ can be a first indicator of the $^{56}$Ni mass produced. It is, then, informative to compare $M^\mathrm{stellar}_\mathrm{ej,>5GK}$ and $M^\mathrm{stellar}_\mathrm{ej,Ni}$. We limit this analysis to the stellar component of the ejecta as it is the matter for which we know the complete thermodynamical history.

This comparison is shown in Fig.~\ref{fig:M5GK_vs_MNi_only_our_data} where the two quantities are displayed as a function of the explosion energy. In most of the models $M^\mathrm{stellar}_\mathrm{ej,>5GK}$ is a good approximation of $M^\mathrm{stellar}_\mathrm{ej,Ni}$. For high energy ($E_\mathrm{expl}\gtrsim 1.5\times10^{51}$~erg) the amount of matter experiencing T$>$5GK is larger than the $^{56}$Ni mass, however, is not always the case, especially at lower energy ($E_\mathrm{expl}\lesssim 0.25\times10^{51}$~erg) when for many explosions we measure $M^\mathrm{stellar}_\mathrm{ej,>5GK}<M^\mathrm{stellar}_\mathrm{ej,Ni}$, which means that $^{56}$Ni starts being produced already at $T<5$GK. 

The results of the $^{56}$Ni mass are expected to be different if the wind is composed of lower electron fraction matter (i.e., $Y_e \ll 0.5$). In this case the nucleosynthesis result is not supposed to peak at $^{56}$Ni, but in heavier nuclei (\citealt{Siegel_2019Natur}) and the amount of $^{56}$Ni produced in the injected component is not expected to be significant. Therefore the $^{56}$Ni mass is not supposed to be dominated by  $M_\mathrm{ej}^\mathrm{inj}$.

\section{Discussion}\label{sec:Discussion}

\subsection{Variety of disk wind-driven explosions}
The progenitor model used in this work is, as mentioned in Sec.~\ref{sec:Method}, we a rapidly rotating, low metallicity, rotationally mixed quasi-chemically homogeneously evolving star with the zero-age main-sequence mass $M_\mathrm{ZAMS}=20\, M_\odot$ taken from \citet{Aguilera-Dena_2020}, which has a very compact core, and it is supposed to fail the explosion (\citealt{Ertl_2016} and \citealt{Muller_2016}).
Our results show a large variety of explosion energies with $E_\mathrm{expl}$ ranging from $\sim 0.049\times10^{51}$~erg to $\sim 34\times10^{51}$~erg (see also Table~\ref{tab:models}).
Moreover, the distribution of these points in the plots can be divided into two categories: the sub-energetic explosions and the highly-energetic ones.
In the plots in Fig.~\ref{fig:Mej_vej_Eexpl}, every single point is the result obtained for the same structure of the progenitor in which we varied the parameters controlling the wind injection. It means that, in reality, a given structure of the progenitor results in only a single point in these plots.
This means that, in reality, the $20M_\odot$ star we employed results in either a highly- or sub-energetic explosion depending on the power of the wind injection and the competition between the ram pressure of the wind and the infalling envelope. 
%\sho{Emphasize that a given structure of the progenitor results in only a single point in, e.g., Fig. 5. Therefore, in reality, the $20M_\odot$ star we employed results in either a highly- or sub-energetic explosion depending on the power of the wind injection.} 

It can be, then, interesting to study this competition for other massive progenitors with different structures to further investigate the variety of explosion properties and to verify if stars with other structures also present sub- and highly-energetic explosion branches depending on the power of the wind injection. The first parameter worthy of consideration is the progenitor mass. For instance, generally speaking, more massive stars tend to present a more compact CO-layer, and thus, have higher mass-infall rates leading to a larger amount of matter available to form the disk and to a higher thermal energy budget for the explosion energy. Therefore, employing such progenitors could allow us to model even more energetic SNe. 

Another parameter that affects the fate of the core collapse and plays a role in the explosion after the formation of a BH is the rotation. The initial angular momentum profile of the progenitor has an impact on the explosion energy and the ejecta mass as shown by \citet{Fujibayashi2023arxiv}. Specifically, a star with fast rotation is expected to determine a more energetic explosion and to enhance the mass ejection. Then, we can also assume more $^{56}$Ni will be produced. Therefore it can be informative also to perform simulations employing the same progenitor star but using different rotational profiles.

In order to investigate the variety of explosion properties depending on the progenitor structure and wind injection, it can be also interesting to use our central engine model for a failed CCSN for those progenitors that are more likely to undergo a neutrino-driven explosion according to \citet{Ertl_2016} or \citet{Muller_2016}. Comparing the results obtained using our code for these two categories of progenitors (those supposed to fail the explosion and those expected to succeed in a neutrino-driven explosion) can be useful also to investigate the different dependencies of the explosion. Indeed \citet{Fujibayashi2023arxiv} showed that the explosion in failed CCSN is sensitive to the mass-infall rate at the disk formation, i.e., a higher mass-infall rate (usually from the carbon-oxygen layer of the star) enhances the viscous and shock heating rates around the inner region of the disk, which result in the larger explosion energy, while \citet{Ertl_2016} found that the neutrino-driven explosion is more likely to be sensitive to the compactness of inner domain, i.e. the iron-silicon layer. Hence, using the same general central engine model can allow us to verify these effects.

As mentioned in Sec.~\ref{sec:parameter_dependence}, another interesting result of our study is that the correlation between $E_\mathrm{expl}$ and $M_\mathrm{ej}$ found in our models is stronger than that of observations (Fig.~\ref{fig:Eexp_vs_Mej_tw_parameter_dependence}).
This illustrates the incapability of our simulations to fully explain the variety of observational data by utilizing a single progenitor model.
However, changing the mass and the rotational profile of the progenitor may fill the areas in Fig.~\ref{fig:Eexp_vs_Mej_tw_parameter_dependence} not covered by the present results.
Performing explosion simulations using progenitors with different $M_\mathrm{ZAMS}$ and rotational profiles is, thus, demanded to further investigate the correlation between the explosion energy and the ejecta mass.

\subsection{The model of the wind injection}
In this work, we adopt a simple prescription to set the wind by evolving the disk and the BH using Eqs.~\eqref{M_dot_disk}-\eqref{J_dot_BH}. In this model, the wind time scale and the accretion time scale are free parameters, which remain constant throughout the simulation.
%Varying $t_\mathrm{w}$ allows us not to be limited to a specific scenario of the wind injection. 
The flexibility in setting the parameters allows us to set engine models with diverse characteristics, unrestricted by specific scenarios.
In Sec.~\ref{sec:parameter_dependence} and \ref{sec:Ni_production}, we showed that the results of our models with $t_\mathrm{w}\sim t_\mathrm{acc}\gtrsim 3.16$~s reproduce $M_\mathrm{ej}$ and $E_\mathrm{expl}$ obtained by a general relativistic neutrino-radiation viscous-hydrodynamics simulation by \citet{Fujibayashi2023arxiv} (see Fig.~\ref{fig:Eexp_vs_Mej_tw_parameter_dependence}).
Our models can also account for the $^{56}$Ni mass measured in their work within the uncertainty of the mass fraction of $^{56}$Ni of the injected component.
These similar results can be explained by the similar time scales in the two simulations at the onset of the injection (i.e., $t \lesssim 20$~s). In the model used by \citet{Fujibayashi2023arxiv} $t_\mathrm{w}$ and $t_\mathrm{acc}$ later evolve in time since they are evaluated from the viscous time scale that depends on the disk radius. However, this happens after the wind injection and they found that at early time in their $20\,M_\odot$ model both $t_\mathrm{w}$ and $t_\mathrm{acc}$ are of several seconds, hence they are comparable to those used in our simulations with $t_\mathrm{w}\sim t_\mathrm{acc}\gtrsim 3.16$~s.

\citet{Fujibayashi2023arxiv} showed that in the viscosity-driven wind scenario, the viscous heating plays a role in determining the injected energy budget.
After the efficiency of neutrino cooling drops, the outflow from the disk is driven by viscosity determining the evolution of the viscous time scale and hence of both $t_\mathrm{w}$ and $t_\mathrm{acc}$.
Therefore, focusing on this scenario, a future implementation of our model can be taking into account the effects of the time evolution of $t_\mathrm{w}$ and $t_\mathrm{acc}$. 
This will allow us to provide more quantitative models for more sophisticated, although specific, scenarios in which we can describe the accretion flow during both the neutrino dominated accretion flow phase (NDAF) (\citealt{Narayan_Yi_1994}; \citealt{Popham_Woosley_1999}) and the advection dominated accretion flow phase (ADAF) (\citealt{Narayan_Yi_1994}, \citealt{Kohri_2005} and \citealt{Hayakawa_2018}). 

\subsection{The effect of GRB ejecta}
As mentioned in Sec.~\ref{sec:Method}, we exclude the central engine from the computational domain, and it is considered as being embedded in the central part of the star and characterized by the presence of a BH and a disk evolved according to Eqs.~\eqref{M_dot_disk}-\eqref{J_dot_BH}.

If the dimensionless spin of the BH is large, in the presence of electromagnetic fields, the Blandford-Znajek effect could play an important role (see \citealt{Blandford_Znajek}), i.e., it could launch an energetic jet or outflow along the spin axis of the BH.
If a relativistic jet is produced, a gamma-ray burst will be also launched (see \citealt{Izzard_Robert_Tout} and \citealt{Gottlieb_Ore_Lalakos} for simulation works). In the presence of the jet, more energy can be injected into the stellar matter, and hence, the energy budget available for the explosion and the $^{56}$Ni production increases. Therefore performing relativistic-hydrodynamic simulations including the injection of relativistic jets will be one of our follow-up works.

\section{Summary and conclusions}\label{sec:Conclusion}
We studied the hydrodynamics and nucleosynthesis for the explosion of a massive star to explore the properties of ejecta and the $^{56}$Ni production in the collapsar scenario. 
Our main goal was to investigate the explosion mechanism of Type Ic SNe in the collapsar scenario.

We implemented a new feature that solves the cylindrically symmetric gravitational potential $\Phi$ to the open-source multi-dimensional hydrodynamics code \texttt{Athena++}.
We used it to simulate the explosion of a rapidly rotating, rotationally mixed, quasi-chemically homogeneous $M_\mathrm{ZAMS}\sim 20\,M_\odot$ star employing the progenitor model from \citet{Aguilera-Dena_2020}. For this work we also built a semi-analytical model for the central engine by taking into account the BH and disk evolution connected through matter and angular momentum transfer, to which we added the contribution of the disk wind following \citet{Hayakawa_2018}.

We tested different models by varying the parameters that control the properties of mass and energy injection to thoroughly investigate their influence on the final ejecta.
The parameter setups of the simulations and the main results are listed in Table~\ref{tab:models}.

In all of our models, we found that the energy and mass injection occurs roughly between 10 and 20~s after the disk formation. After the wind injection, the competition between the ram pressure of the injected and infalling matter leads the disk wind-driven explosion to be sub- or highly-energetic with an explosion energy $<\SI{0.1e51}{erg}$ and $>\SI{e52}{erg}$ respectively.
 
This distinction originates from whether the $P_\mathrm{ram}$ of the injected matter  can overcome $P_\mathrm{ram}$ of the infalling envelope and efficiently push the stellar envelope outwards.
When the first one is larger than the second one, most of the matter can expand outwards without falling back, leading to a highly-energetic explosion with $\sim \SI{1e52}{erg}$.
In the case of the sub-energetic explosions ($E_\mathrm{expl}<\SI{1e51}{erg}$), instead, the shock wave transfers the energy from the infalling matter bouncing on the wind. Propagating then outwards, the bounce shock causes the ejection of the matter outside the star.

Studying the impact of the parameters on the final ejecta, we found that the wind timescale strongly affects $E_\mathrm{expl}$. In particular, models with longer wind time scales tend to reach higher explosion energies due to a high mass-infall rate.  
We also noticed that models with higher $\xi^2$ (i.e. $\xi^2=0.3$), and hence, higher wind kinetic energy, or smaller $f_\mathrm{therm}$ (that is $f_\mathrm{therm}=0.01$) represent highly-energetic explosions. In contrast, smaller values of $\xi^2$ or larger $f_\mathrm{therm}$ lead the explosions to be sub-energetic.

We found that sub-energetic explosions produce smaller amounts of $^{56}$Ni  ($<0.2\,M_\odot$) while highly-energetic explosions produce $0.2-2.1 \,M_\odot$ of $^{56}$Ni. 
The $^{56}$Ni mass was evaluated separately taking into account the stellar and the injected components of the ejecta since the whole thermodynamical history of the particle tracer is only available for the former component.
Our results show that $M^\mathrm{inj}_\mathrm{ej}$ dominates and the difference between $M^\mathrm{stellar}_\mathrm{ej, Ni}$ and $M^\mathrm{stellar}_\mathrm{ej, Ni}+M^\mathrm{inj}_\mathrm{ej}$ is larger for highly-energetic explosions.

We also compared our numerical results with the observational data for stripped-envelope SNe, some of which are broad-lined type Ic SNe, taken from \citet{Taddia2019jan} and \citet{Gomez2022dec}.
We found that the distribution of the $^{56}$Ni mass of our models reproduces the relation between $M_\mathrm{Ni}$ and $E_\mathrm{expl}$ or $M_\mathrm{Ni}$ and $v_\mathrm{ej}$ for very-high energy SNe with $E_\mathrm{expl}>2\times 10^{51}$ erg and sub-energetic SNe with $E_\mathrm{expl}< 0.1\times10^{51}$ erg of the observational data.
Moreover, we measured a tighter correlation between the explosion energy and the ejecta mass in our simulations than that observationally measured by \citet{Taddia2019jan}  and \citet{Gomez2022dec}. 

To better investigate the variety of explosion properties and to verify whether stars with different structures present sub- and highly-energetic explosion branches, we plan to perform numerical simulations by varying the mass and the rotational profile of the progenitor in our follow-up work. 
More massive stars can, for instance, have a larger amount of matter to form the disk and hence a higher energy budget for the explosion energy. 
Moreover, \citet{Fujibayashi2023jan} showed that the initial angular momentum profile of the progenitor affects the explosion energy and the ejecta mass.
Therefore we expect that varying the progenitor mass and rotational profile may explain the variety of observational data.

Finally, in this work, we found that our models with $t_\mathrm{w}\sim t_\mathrm{acc}$ of several seconds reproduce the results for $M_\mathrm{ej}$, $E_\mathrm{expl}$ and the $^{56}$Ni mass obtained by \citet{Fujibayashi2023arxiv} in a general relativistic neutrino-radiation viscous-hydrodynamics simulation that utilizes the same progenitor model. 
This is due to the similar time scales in our and their works at early times (i.e., $t<20$s), at the onset of the injection. In their model for the viscosity-driven wind scenario, however, $t_\mathrm{w}$ and $t_\mathrm{acc}$  later evolve in time with the viscous time scale. However this happens after the wind injection which is the moment that we model in this work and we are focusing on. At this early time $t_\mathrm{w}$ and $t_\mathrm{acc}$ in our work and that of \citet{Fujibayashi2023arxiv} are similar. Considering the model used by \citet{Fujibayashi2023arxiv}, it may be interesting implement the evolution of these time scales in our model and use it to investigate the NDAF and ADAF phases in this specific scenario.

\section*{Acknowledgements}
We want to thank David Aguilera-Dena for providing his stellar evolution model and Kengo Tomida for his helpful guide in using \texttt{Athena++}. We are thankful for the useful and constructive discussion we had with Alexis Reboul-Salze, Aurore Betranhandy, Dina Traykova, Kyohei Kawaguchi and  Johannes Ringler during the revision of this paper. We also want to thank the anonymous referee for the comments that improved the manuscript.
This study was supported in part by Grants-in-Aid for Scientific Research of the Japan Society for the Promotion of Science (JSPS, No. JP22K20377).
Numerical computation was performed on Sakura cluster at Max Planck Computing and Data Facility.

\section*{Data availability}
The data will be shared on reasonable request to the corresponding author.


\bibliographystyle{mnras}
\bibliography{biblio}

\begin{thebibliography}{}
\makeatletter
\relax
\def\mn@urlcharsother{\let\do\@makeother \do\$\do\&\do\#\do\^\do\_\do\%\do\~}
\def\mn@doi{\begingroup\mn@urlcharsother \@ifnextchar [ {\mn@doi@} {\mn@doi@[]}}
\def\mn@doi@[#1]#2{\def\@tempa{#1}\ifx\@tempa\@empty \href {http://dx.doi.org/#2} {doi:#2}\else \href {http://dx.doi.org/#2} {#1}\fi \endgroup}
\def\mn@eprint#1#2{\mn@eprint@#1:#2::\@nil}
\def\mn@eprint@arXiv#1{\href {http://arxiv.org/abs/#1} {{\tt arXiv:#1}}}
\def\mn@eprint@dblp#1{\href {http://dblp.uni-trier.de/rec/bibtex/#1.xml} {dblp:#1}}
\def\mn@eprint@#1:#2:#3:#4\@nil{\def\@tempa {#1}\def\@tempb {#2}\def\@tempc {#3}\ifx \@tempc \@empty \let \@tempc \@tempb \let \@tempb \@tempa \fi \ifx \@tempb \@empty \def\@tempb {arXiv}\fi \@ifundefined {mn@eprint@\@tempb}{\@tempb:\@tempc}{\expandafter \expandafter \csname mn@eprint@\@tempb\endcsname \expandafter{\@tempc}}}

\bibitem[\protect\citeauthoryear{Aguilera-Dena, Langer, Antoniadis  \& Müller}{Aguilera-Dena et~al.}{2020}]{Aguilera-Dena_2020}
Aguilera-Dena D.~R.,  Langer N.,  Antoniadis J.,   Müller B.,  2020, \mn@doi [The Astrophysical Journal] {10.3847/1538-4357/abb138}, 901, 114

\bibitem[\protect\citeauthoryear{{Aloy} \& {Obergaulinger}}{{Aloy} \& {Obergaulinger}}{2021}]{Obergaulinger_Aloy_2021_1}
{Aloy} M.~{\'A}.,  {Obergaulinger} M.,  2021, \mn@doi [\mnras] {10.1093/mnras/staa3273}, \href {https://ui.adsabs.harvard.edu/abs/2021MNRAS.500.4365A} {500, 4365}

\bibitem[\protect\citeauthoryear{{Balbus} \& {Hawley}}{{Balbus} \& {Hawley}}{1991}]{Balbus1991a}
{Balbus} S.~A.,  {Hawley} J.~F.,  1991, \mn@doi [\apj] {10.1086/170270}, \href {https://ui.adsabs.harvard.edu/abs/1991ApJ...376..214B} {376, 214}

\bibitem[\protect\citeauthoryear{Balbus \& Hawley}{Balbus \& Hawley}{1998}]{Balbus:1998ja}
Balbus S.~A.,  Hawley J.~F.,  1998, \mn@doi [Rev. Mod. Phys.] {10.1103/RevModPhys.70.1}, 70, 1

\bibitem[\protect\citeauthoryear{{Bardeen}, {Press}  \& {Teukolsky}}{{Bardeen} et~al.}{1972}]{Bardeen_ISCO}
{Bardeen} J.~M.,  {Press} W.~H.,   {Teukolsky} S.~A.,  1972, \mn@doi [\apj] {10.1086/151796}, \href {https://ui.adsabs.harvard.edu/abs/1972ApJ...178..347B} {178, 347}

\bibitem[\protect\citeauthoryear{{Blandford} \& {Payne}}{{Blandford} \& {Payne}}{1982}]{Blandford1982jun}
{Blandford} R.~D.,  {Payne} D.~G.,  1982, \mn@doi [\mnras] {10.1093/mnras/199.4.883}, \href {https://ui.adsabs.harvard.edu/abs/1982MNRAS.199..883B} {199, 883}

\bibitem[\protect\citeauthoryear{Blandford \& Znajek}{Blandford \& Znajek}{1977}]{Blandford_Znajek}
Blandford R.~D.,  Znajek R.~L.,  1977, \mn@doi [Monthly Notices of the Royal Astronomical Society] {10.1093/mnras/179.3.433}, 179, 433

\bibitem[\protect\citeauthoryear{Burrows, Radice, Vartanyan, Nagakura, Skinner  \& Dolence}{Burrows et~al.}{2019}]{Burrows_2020}
Burrows A.,  Radice D.,  Vartanyan D.,  Nagakura H.,  Skinner M.~A.,   Dolence J.~C.,  2019, \mn@doi [Monthly Notices of the Royal Astronomical Society] {10.1093/mnras/stz3223}, 491, 2715

\bibitem[\protect\citeauthoryear{{Ertl}, {Janka}, {Woosley}, {Sukhbold}  \& {Ugliano}}{{Ertl} et~al.}{2016}]{Ertl_2016}
{Ertl} T.,  {Janka} H.~T.,  {Woosley} S.~E.,  {Sukhbold} T.,   {Ugliano} M.,  2016, \mn@doi [\apj] {10.3847/0004-637X/818/2/124}, \href {https://ui.adsabs.harvard.edu/abs/2016ApJ...818..124E} {818, 124}

\bibitem[\protect\citeauthoryear{{Frail} et~al.,}{{Frail} et~al.}{2001}]{2001_Frail}
{Frail} D.~A.,  et~al., 2001, \mn@doi [\apjl] {10.1086/338119}, \href {https://ui.adsabs.harvard.edu/abs/2001ApJ...562L..55F} {562, L55}

\bibitem[\protect\citeauthoryear{{Fujibayashi}, {Tsz-Lok Lam}, {Shibata}  \& {Sekiguchi}}{{Fujibayashi} et~al.}{2023a}]{Fujibayashi2023arxiv}
{Fujibayashi} S.,  {Tsz-Lok Lam} A.,  {Shibata} M.,   {Sekiguchi} Y.,  2023a, \mn@doi [arXiv e-prints] {10.48550/arXiv.2309.02161}, \href {https://ui.adsabs.harvard.edu/abs/2023arXiv230902161F} {p. arXiv:2309.02161}

\bibitem[\protect\citeauthoryear{{Fujibayashi}, {Kiuchi}, {Wanajo}, {Kyutoku}, {Sekiguchi}  \& {Shibata}}{{Fujibayashi} et~al.}{2023b}]{Fujibayashi2023jan}
{Fujibayashi} S.,  {Kiuchi} K.,  {Wanajo} S.,  {Kyutoku} K.,  {Sekiguchi} Y.,   {Shibata} M.,  2023b, \mn@doi [\apj] {10.3847/1538-4357/ac9ce0}, \href {https://ui.adsabs.harvard.edu/abs/2023ApJ...942...39F} {942, 39}

\bibitem[\protect\citeauthoryear{{Gomez}, {Berger}, {Nicholl}, {Blanchard}  \& {Hosseinzadeh}}{{Gomez} et~al.}{2022}]{Gomez2022dec}
{Gomez} S.,  {Berger} E.,  {Nicholl} M.,  {Blanchard} P.~K.,   {Hosseinzadeh} G.,  2022, \mn@doi [\apj] {10.3847/1538-4357/ac9842}, \href {https://ui.adsabs.harvard.edu/abs/2022ApJ...941..107G} {941, 107}

\bibitem[\protect\citeauthoryear{Gottlieb, Lalakos, Bromberg, Liska  \& Tchekhovskoy}{Gottlieb et~al.}{2022}]{Gottlieb_Ore_Lalakos}
Gottlieb O.,  Lalakos A.,  Bromberg O.,  Liska M.,   Tchekhovskoy A.,  2022, \mn@doi [Monthly Notices of the Royal Astronomical Society] {10.1093/mnras/stab3784}, 510, 4962

\bibitem[\protect\citeauthoryear{{Grimmett}, {M{\"u}ller}, {Heger}, {Banerjee}  \& {Obergaulinger}}{{Grimmett} et~al.}{2021}]{Grimmet_HNe_2021}
{Grimmett} J.~J.,  {M{\"u}ller} B.,  {Heger} A.,  {Banerjee} P.,   {Obergaulinger} M.,  2021, \mn@doi [\mnras] {10.1093/mnras/staa3819}, \href {https://ui.adsabs.harvard.edu/abs/2021MNRAS.501.2764G} {501, 2764}

\bibitem[\protect\citeauthoryear{{Hachisu}}{{Hachisu}}{1986}]{Hachisu_gravitysolver}
{Hachisu} I.,  1986, \mn@doi [\apjs] {10.1086/191148}, \href {https://ui.adsabs.harvard.edu/abs/1986ApJS...62..461H} {62, 461}

\bibitem[\protect\citeauthoryear{Hayakawa \& Maeda}{Hayakawa \& Maeda}{2018}]{Hayakawa_2018}
Hayakawa T.,  Maeda K.,  2018, \mn@doi [The Astrophysical Journal] {10.3847/1538-4357/aaa76c}, 854, 43

\bibitem[\protect\citeauthoryear{Izzard, Tout, Karakas  \& Pols}{Izzard et~al.}{2004}]{Izzard_Robert_Tout}
Izzard R.~G.,  Tout C.~A.,  Karakas A.~I.,   Pols O.~R.,  2004, \mn@doi [Monthly Notices of the Royal Astronomical Society] {10.1111/j.1365-2966.2004.07446.x}, 350, 407

\bibitem[\protect\citeauthoryear{{Janiuk, Agnieszka}, {Charzy\'{}nski, Szymon}  \& {Bejger, Michal}}{{Janiuk, Agnieszka} et~al.}{2013}]{Janiuk_2013}
{Janiuk, Agnieszka} {Charzy\'{}nski, Szymon}  {Bejger, Michal} 2013, \mn@doi [A\&A] {10.1051/0004-6361/201322165}, 560, A25

\bibitem[\protect\citeauthoryear{{Japelj}, {Vergani}, {Salvaterra}, {Hunt}  \& {Mannucci}}{{Japelj} et~al.}{2016}]{Japelj_2016}
{Japelj} J.,  {Vergani} S.~D.,  {Salvaterra} R.,  {Hunt} L.~K.,   {Mannucci} F.,  2016, \mn@doi [\aap] {10.1051/0004-6361/201628603}, \href {https://ui.adsabs.harvard.edu/abs/2016A&A...593A.115J} {593, A115}

\bibitem[\protect\citeauthoryear{{Just}, {Aloy}, {Obergaulinger}  \& {Nagataki}}{{Just} et~al.}{2022}]{Just2022}
{Just} O.,  {Aloy} M.~A.,  {Obergaulinger} M.,   {Nagataki} S.,  2022, \mn@doi [\apjl] {10.3847/2041-8213/ac83a1}, \href {https://ui.adsabs.harvard.edu/abs/2022ApJ...934L..30J} {934, L30}

\bibitem[\protect\citeauthoryear{Kohri, Narayan  \& Piran}{Kohri et~al.}{2005}]{Kohri_2005}
Kohri K.,  Narayan R.,   Piran T.,  2005, \mn@doi [The Astrophysical Journal] {10.1086/431354}, 629, 341

\bibitem[\protect\citeauthoryear{Kumar, Narayan  \& Johnson}{Kumar et~al.}{2008}]{Kumar_Narayan_method}
Kumar P.,  Narayan R.,   Johnson J.~L.,  2008, \mn@doi [Monthly Notices of the Royal Astronomical Society] {10.1111/j.1365-2966.2008.13493.x}, 388, 1729

\bibitem[\protect\citeauthoryear{Liu, Song, Yi, Gu  \& Wang}{Liu et~al.}{2019}]{LIU20195}
Liu T.,  Song C.-Y.,  Yi T.,  Gu W.-M.,   Wang X.-F.,  2019, \mn@doi [Journal of High Energy Astrophysics] {https://doi.org/10.1016/j.jheap.2019.02.001}, 22, 5

\bibitem[\protect\citeauthoryear{MacFadyen \& Woosley}{MacFadyen \& Woosley}{1999}]{macfadyen1999collapsars}
MacFadyen A.,  Woosley S.,  1999, The Astrophysical Journal, 524, 262

\bibitem[\protect\citeauthoryear{Margalit, Metzger, Thompson, Nicholl  \& Sukhbold}{Margalit et~al.}{2018}]{Margalit_2018}
Margalit B.,  Metzger B.~D.,  Thompson T.~A.,  Nicholl M.,   Sukhbold T.,  2018, \mn@doi [Monthly Notices of the Royal Astronomical Society] {10.1093/mnras/sty013}, 475, 2659

\bibitem[\protect\citeauthoryear{M{\'{e}}sz{\'{a}}ros}{M{\'{e}}sz{\'{a}}ros}{2006}]{Meszaros_2006}
M{\'{e}}sz{\'{a}}ros P.,  2006, \mn@doi [Reports on Progress in Physics] {10.1088/0034-4885/69/8/r01}, 69, 2259

\bibitem[\protect\citeauthoryear{{M{\"u}ller}, {Heger}, {Liptai}  \& {Cameron}}{{M{\"u}ller} et~al.}{2016}]{Muller_2016}
{M{\"u}ller} B.,  {Heger} A.,  {Liptai} D.,   {Cameron} J.~B.,  2016, \mn@doi [\mnras] {10.1093/mnras/stw1083}, \href {https://ui.adsabs.harvard.edu/abs/2016MNRAS.460..742M} {460, 742}

\bibitem[\protect\citeauthoryear{{Narayan} \& {Yi}}{{Narayan} \& {Yi}}{1994}]{Narayan_Yi_1994}
{Narayan} R.,  {Yi} I.,  1994, \mn@doi [\apjl] {10.1086/187381}, \href {https://ui.adsabs.harvard.edu/abs/1994ApJ...428L..13N} {428, L13}

\bibitem[\protect\citeauthoryear{{O'Connor} \& {Ott}}{{O'Connor} \& {Ott}}{2011}]{OConnor2011}
{O'Connor} E.,  {Ott} C.~D.,  2011, \mn@doi [\apj] {10.1088/0004-637X/730/2/70}, \href {https://ui.adsabs.harvard.edu/abs/2011ApJ...730...70O} {730, 70}

\bibitem[\protect\citeauthoryear{{Obergaulinger} \& {Aloy}}{{Obergaulinger} \& {Aloy}}{2017}]{Obergaulinger_Aloy_2017}
{Obergaulinger} M.,  {Aloy} M.~{\'A}.,  2017, \mn@doi [\mnras] {10.1093/mnrasl/slx046}, \href {https://ui.adsabs.harvard.edu/abs/2017MNRAS.469L..43O} {469, L43}

\bibitem[\protect\citeauthoryear{{Obergaulinger} \& {Aloy}}{{Obergaulinger} \& {Aloy}}{2020}]{Obergaulinger_Aloy_2020}
{Obergaulinger} M.,  {Aloy} M.~{\'A}.,  2020, \mn@doi [\mnras] {10.1093/mnras/staa096}, \href {https://ui.adsabs.harvard.edu/abs/2020MNRAS.492.4613O} {492, 4613}

\bibitem[\protect\citeauthoryear{{Obergaulinger} \& {Aloy}}{{Obergaulinger} \& {Aloy}}{2021}]{Obergaulinger_Aloy_2021_2}
{Obergaulinger} M.,  {Aloy} M.~{\'A}.,  2021, \mn@doi [\mnras] {10.1093/mnras/stab295}, \href {https://ui.adsabs.harvard.edu/abs/2021MNRAS.503.4942O} {503, 4942}

\bibitem[\protect\citeauthoryear{Obergaulinger \& Aloy}{Obergaulinger \& Aloy}{2022}]{10.1093/mnras/stac613}
Obergaulinger M.,  Aloy M.~A.,  2022, \mn@doi [Monthly Notices of the Royal Astronomical Society] {10.1093/mnras/stac613}, 512, 2489

\bibitem[\protect\citeauthoryear{{Obergaulinger}, {Aloy}  \& {M{\"u}ller}}{{Obergaulinger} et~al.}{2010}]{Obergaulinger2010jun}
{Obergaulinger} M.,  {Aloy} M.~A.,   {M{\"u}ller} E.,  2010, \mn@doi [\aap] {10.1051/0004-6361/200913386}, \href {https://ui.adsabs.harvard.edu/abs/2010A&A...515A..30O} {515, A30}

\bibitem[\protect\citeauthoryear{Papish \& Soker}{Papish \& Soker}{2013}]{10.1093/mnras/stt2199}
Papish O.,  Soker N.,  2013, Monthly Notices of the Royal Astronomical Society, 438, 1027

\bibitem[\protect\citeauthoryear{{Piran}}{{Piran}}{2004}]{2004_Piran}
{Piran} T.,  2004, \mn@doi [Reviews of Modern Physics] {10.1103/RevModPhys.76.1143}, \href {https://ui.adsabs.harvard.edu/abs/2004RvMP...76.1143P} {76, 1143}

\bibitem[\protect\citeauthoryear{{Popham}, {Woosley}  \& {Fryer}}{{Popham} et~al.}{1999}]{Popham_Woosley_1999}
{Popham} R.,  {Woosley} S.~E.,   {Fryer} C.,  1999, \mn@doi [\apj] {10.1086/307259}, \href {https://ui.adsabs.harvard.edu/abs/1999ApJ...518..356P} {518, 356}

\bibitem[\protect\citeauthoryear{{Siegel}, {Barnes}  \& {Metzger}}{{Siegel} et~al.}{2019}]{Siegel_2019Natur}
{Siegel} D.~M.,  {Barnes} J.,   {Metzger} B.~D.,  2019, \mn@doi [\nat] {10.1038/s41586-019-1136-0}, \href {https://ui.adsabs.harvard.edu/abs/2019Natur.569..241S} {569, 241}

\bibitem[\protect\citeauthoryear{{Sieverding}, {Waldrop}, {Harris}, {Hix}, {Lentz}, {Bruenn}  \& {Messer}}{{Sieverding} et~al.}{2023}]{Sieverding2023jun}
{Sieverding} A.,  {Waldrop} P.~G.,  {Harris} J.~A.,  {Hix} W.~R.,  {Lentz} E.~J.,  {Bruenn} S.~W.,   {Messer} O.~E.~B.,  2023, \mn@doi [\apj] {10.3847/1538-4357/acc8d1}, \href {https://ui.adsabs.harvard.edu/abs/2023ApJ...950...34S} {950, 34}

\bibitem[\protect\citeauthoryear{Stone, Tomida, White  \& Felker}{Stone et~al.}{2020}]{Stone_2020}
Stone J.~M.,  Tomida K.,  White C.~J.,   Felker K.~G.,  2020, \mn@doi [\apjs] {10.3847/1538-4365/ab929b}, 249, 4

\bibitem[\protect\citeauthoryear{{Sukhbold} \& {Woosley}}{{Sukhbold} \& {Woosley}}{2014}]{Sukhbold_2014}
{Sukhbold} T.,  {Woosley} S.~E.,  2014, \mn@doi [\apj] {10.1088/0004-637X/783/1/10}, \href {https://ui.adsabs.harvard.edu/abs/2014ApJ...783...10S} {783, 10}

\bibitem[\protect\citeauthoryear{{Taddia} et~al.,}{{Taddia} et~al.}{2019a}]{Taddia_2019}
{Taddia} F.,  et~al., 2019a, \mn@doi [\aap] {10.1051/0004-6361/201834429}, \href {https://ui.adsabs.harvard.edu/abs/2019A&A...621A..71T} {621, A71}

\bibitem[\protect\citeauthoryear{{Taddia} et~al.,}{{Taddia} et~al.}{2019b}]{Taddia2019jan}
{Taddia} F.,  et~al., 2019b, \mn@doi [\aap] {10.1051/0004-6361/201834429}, \href {https://ui.adsabs.harvard.edu/abs/2019A&A...621A..71T} {621, A71}

\bibitem[\protect\citeauthoryear{{Timmes}, {Hoffman}  \& {Woosley}}{{Timmes} et~al.}{2000}]{Timmes2000jul}
{Timmes} F.~X.,  {Hoffman} R.~D.,   {Woosley} S.~E.,  2000, \mn@doi [\apjs] {10.1086/313407}, \href {https://ui.adsabs.harvard.edu/abs/2000ApJS..129..377T} {129, 377}

\bibitem[\protect\citeauthoryear{Trigo-Rodr{\'\i}guez, Cano, Wang, Dai  \& Wu}{Trigo-Rodr{\'\i}guez et~al.}{2017}]{Rodriguez_Cano}
Trigo-Rodr{\'\i}guez J.~M.,  Cano Z.,  Wang S.-Q.,  Dai Z.-G.,   Wu X.-F.,  2017, The Observer's Guide to the Gamma-Ray Burst Supernova Connection, \mn@doi{10.1155/2017/8929054.
}, \url {https://doi.org/10.1155/2017/8929054}

\bibitem[\protect\citeauthoryear{Wang \& Burrows}{Wang \& Burrows}{2023}]{Wang_2023}
Wang T.,  Burrows A.,  2023, \mn@doi [The Astrophysical Journal] {10.3847/1538-4357/ace7b2}, 954, 114

\bibitem[\protect\citeauthoryear{{Woosley}}{{Woosley}}{1993}]{Woosley_1993}
{Woosley} S.~E.,  1993, \mn@doi [\apj] {10.1086/172359}, \href {https://ui.adsabs.harvard.edu/abs/1993ApJ...405..273W} {405, 273}

\bibitem[\protect\citeauthoryear{{Woosley} \& {Bloom}}{{Woosley} \& {Bloom}}{2006}]{Woosley_Bloom_2006}
{Woosley} S.~E.,  {Bloom} J.~S.,  2006, \mn@doi [\araa] {10.1146/annurev.astro.43.072103.150558}, \href {https://ui.adsabs.harvard.edu/abs/2006ARA&A..44..507W} {44, 507}

\bibitem[\protect\citeauthoryear{Woosley \& Heger}{Woosley \& Heger}{2006}]{Woosley_2006}
Woosley S.~E.,  Heger A.,  2006, \mn@doi [The Astrophysical Journal] {10.1086/498500}, 637, 914

\makeatother
\end{thebibliography}

%\begin{thebibliography}{99}
%\end{thebibliography}

%%%%%%%%%%%%%%%%%%%%%%%%%%%%%%%%%%%%%%%%%%%%%%%%%%

%%%%%%%%%%%%%%%%% APPENDICES %%%%%%%%%%%%%%%%%%%%%

\appendix

\section{2D gravity solver}\label{Appendix:2D_gravity_solver}
Hereafter we set $4\pi G=1$ for simplicity.
We can evaluate the gravitational potential in spherical-polar coordinates at $(r,\theta)$ using the multipole expansion as:
\begin{ceqn}
\begin{align}
\Phi(r,\cos{\theta})=&-\int_0^{\infty}{dr'}\int_{-1}^{1}{d\cos{\theta'}}\notag \\ &\times\Bigg(\frac{1}{2}\sum_{n=0}^{\infty}{f_n(r',r)P_n(\cos{\theta})P_n(\cos{\theta'})\rho(r',\cos{\theta'})}\Bigg) \notag \\
=& -\sum_{n}\phi_n(r)P_n(\cos{\theta}), \notag
\end{align}
\end{ceqn}
where $P_n(x)$ is the Legendre's polynomial and it is defined as the coefficients of an expansion of:

\begin{align}
    \frac{1}{\sqrt{1-2xt+t^2}}\equiv \sum_{k=0}^\infty P_n(x)t^k
\end{align}
which converges for $|t|<1$. And $f_n(r',r)$ is given by:
\begin{equation}
f_n(r',r)=
    \begin{cases}
    r'\Big(\frac{r'}{r} \Big)^{n+1}\qquad \,\rm{for}\; r'<r\\
    r'\Big(\frac{r}{r'} \Big)^{n}\qquad \quad\rm{for}\; r'>r
    \end{cases}
\end{equation}
and 
\begin{ceqn}
    \begin{align} \label{phi_n_r}
    \phi_n(r) &= -\frac{1}{2}\int {dr'}\int{d\cos{\theta'} f_n(r',r)P_n(\cos{\theta'})\rho(r',\cos{\theta'})}.
    \end{align}
\end{ceqn}

To express now the potential in the computational domain we consider $i\in [1,I]$ the index for $r_i$ and $j\in [1,J]$ the index for $ \cos\theta_j$ and $N$ be the maximum order of the multipole expansion. In this work $N=5$. Then the equation of the multipole expansion becomes:
\begin{ceqn}
\begin{align}
    \Phi&(i,j)= -\sum_{i'=1}^I\sum_{j'=1}^Jdr_{i'}d\cos{\theta}_{j'}\notag \\
    &\hspace{1cm}\times \Bigg (\frac{1}{2}\sum_{n=0}^{N}f_n(i',i)P_n(j)P_n(j')\rho(i',j') \Bigg) \notag \\
    =& -\sum_{n=0}^N\Bigg(\frac{1}{2}\sum_{i'=1}^I\sum_{j'=1}^J dr_{i'}d\cos{\theta}_{j'}f_n(i',i)P_n(j)Pn(j')\rho(i',j') \Bigg)
\end{align}
\end{ceqn}
where $\phi(i,j)$ is the gravitational potential at $(i,j)$ grid point, $dr_{i'}$ and $d\cos{\theta}_{j'}$ are respectively the $r-$ and $\cos{\theta}-$widths of $(i',j')$ grid point, $\rho(i',j')$ is the density of $(i',j')$ grid point and the coefficients are evaluated as:

\begin{align}
&f_n(i',i)=f_n(r_{i'},r_i)\\
&P_n(j)\;\;\,=P_n(\cos{\theta}_j)\\
&P_n(j')\,\,= P_n(\cos{\theta'}_j)
\end{align}

In this way the summation reduces to:

\begin{align}
\Phi(i,j)=\sum_{n=0}^N P_n(j)\phi_n(i)
\end{align}
with:
\begin{align}
    \phi_n(i)\equiv -\sum_{i'=1}^N\sum_{j'=1}^J \Bigg(\frac{1}{2}dr(i')d\cos{\theta}(j')f_n(i',i)P_n(j')\rho(i'j') \Bigg)
\end{align}
that represents the radial component of the gravitational potential.

\subsection{The contribution to the gravitational potential of the mass outside the computational domain} 
In this appendix we describe the procedure we used to 
If the computational domain is limited between a non-zero inner radius $r_\mathrm{in}$ to an outer radius $r_\mathrm{out}$, then we have to take into account the mass distribution for $r<r_\mathrm{in}$ contributing to the gravitational potential.
Therefore the integral of $\phi_n(r)$ showed in equation~\eqref{phi_n_r} splits in two parts:
\begin{ceqn}
    \begin{align}
        \phi_n(r) =& -\frac{1}{2}\int_0^{r_\mathrm{in}}\int{f_n(r',r)P_n(\cos{\theta'})\rho(r',\cos{\theta'})dr'd\cos{\theta'}}\notag \\
        &-\frac{1}{2}\int_{r_\mathrm{in}}^{r_\mathrm{out}}\int{f_n(r',r)P_n(\cos{\theta'})\rho(r',\cos{\theta'})dr'd\cos{\theta'}}\notag \\
        =&\vcentcolon \phi_n^{<r_\mathrm{in}}(r)+\phi_n^{r_\mathrm{in}<r<r_\mathrm{out}}(r)
    \end{align}
\end{ceqn}

For $r<r_\mathrm{in}$ we only know the total mass of the region and we don't have any information about the denisty distribution there, since it is not part of the computational domain. Therefore we assume a spherically symmetric matter distribution in that central region (with $r<r_\mathrm{in}$): $\rho(r,\cos{\theta}) = \rho(r)$. Because of the assumption of spherical symmetry $\phi_n^{<r_\mathrm{in}}=0$ for $n\ge 0$ and

\begin{align}\label{phi_0_sphsym}
        \phi_0^{<r_\mathrm{in}}&=-\frac{1}{2}\int_0^{r_\mathrm{in}}\int{f_0(r',r)\rho(r') dr'd\cos{\theta'}}\notag \\
        &= -\int_0^{r_\mathrm{in}}{f_0(r',r)\rho(r')dr'}
\end{align}

Since we are interested in computing the potential inside the computational domain in which $r>r_\mathrm{in}$, then $f_0(r',r)$ is always given by $f_0(r',r)=r'^2/r$. Therefore, equation~\eqref{phi_0_sphsym} becomes:
\begin{align}
        \phi_0^{<r_\mathrm{in}}&=-\frac{1}{r}\int_0^{r_\mathrm{in}}{r'^2\rho(r')dr'}\notag \\
        &=-\frac{M^{<r_\mathrm{in}}}{r}.
\end{align}

\section{Model of the disk wind}\label{Appendix:disk_outflow}
If we consider the evolution if the specific angular momentum of the disk $ j_\mathrm{disk}$ as described in equation~\eqref{dJ_disk_over_dM_disk}, we can notice that if the contribution of infalling matter is small, the average specific angular momentum of the disk increases with time. As a consequence the disk radius, which is defined as $r_{\mathrm{disk}}\coloneqq j_\mathrm{disk}^2/(GM_\mathrm{BH})$, can become larger than the inner boundary radius. Indeed if we assume that the wind carries the average specific angular momentum of the disk and we require the angular momentum conservation, considering a uniform angular velocity, $\omega_\mathrm{wind}$ is evaluated as:
\begin{align}
        j_\mathrm{disk}\dot{M}_\mathrm{wind} 
        = 2\pi r^4_\mathrm{in}\omega_\mathrm{w} \int_{\theta_1^*}^{\theta_2^*}{\rho_\mathrm{w}\sin^3\theta d\theta},
\end{align}

Which implies $\omega \sim j_\mathrm{disk}/{r_\mathrm{in}^2}=r_\mathrm{disk}^2/r_\mathrm{in}^2\sqrt{(GM_\mathrm{BH})/r_\mathrm{disk}^3}$.  If $r_\mathrm{disk}/r_\mathrm{in}\ll 1$, then $\omega_\mathrm{wind}$, and hence $v_\phi = r_{\mathrm{in}}\omega$, at the injection can become large, even larger than the speed of light. 

Since our model does not describe the injection from a disk with $r_\mathrm{disk}>r_\mathrm{in}$, we cannot describe such a system consistently. So we describe the disk outflow according to equation~\eqref{e_inj_eqs} considering that the injected matter does not carry angular momentum.

\section{Convergence studies}

\begin{figure}
\centering
	\captionsetup{font=small, labelfont=rm}
	%\subfloat[]
	{\includegraphics [width=0.48\textwidth]{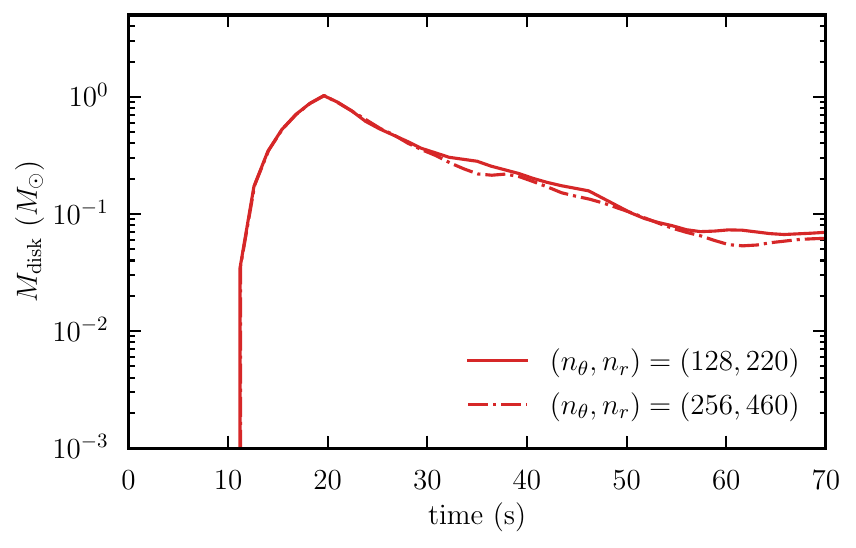}	
		\label{fig:M_disk_higher_res_comp} }\quad
	%\subfloat[]
	{\includegraphics [width=0.48\textwidth]{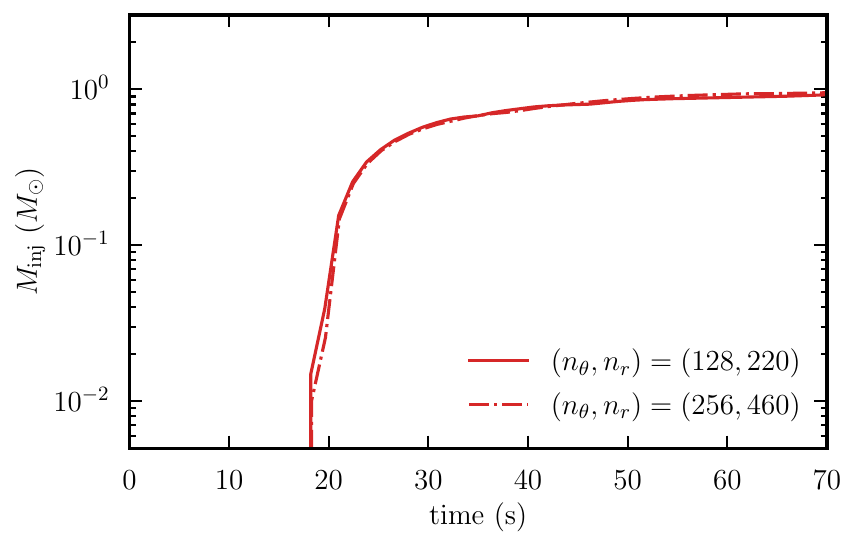}
		\label{fig:M_inj_higher_res_comp} }
\caption{Upper panel: Disk mass of M20\_10\_1\_0.3\_0.1 at two different resolutions. The red solid line shows the results for  $(n_{\theta}, n_{r})=(128,220)$ used in this work and the dotdashed line displays the quantities for $(n_{\theta}, n_{r})=(256,460)$.
Lower panel:injected mass of M20\_10\_1\_0.3\_0.1 at two different resolutions. The  solid line shows the results for $(n_{\theta}, n_{r})=(128,220)$ and the dotdashed line displays the quantities for $(n_{\theta}, n_{r})=(256,460)$.
	}
\label{fig:M_disk_M_inj_higher_res_comp}
\end{figure}

\begin{figure}
	\centering
	\includegraphics [width=0.48\textwidth]{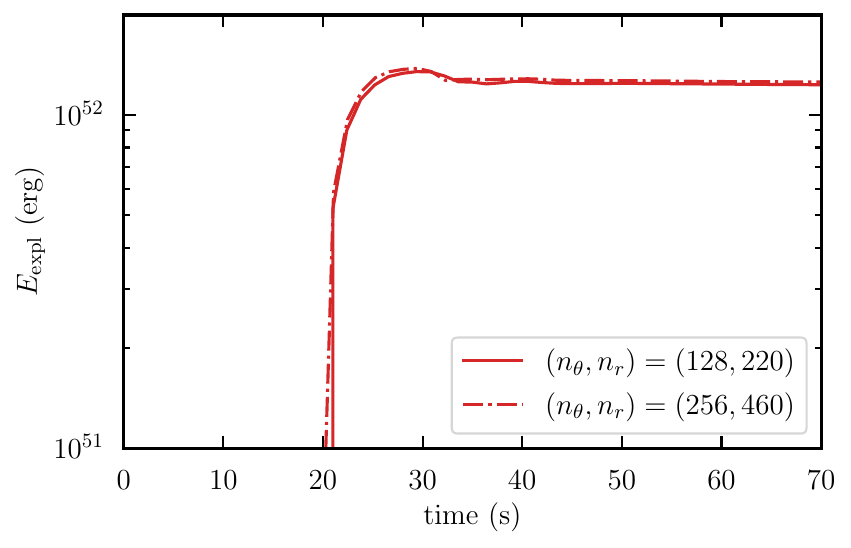}
	\caption{Evolution of the explosion energy of M20\_10\_1\_0.3\_0.1 at two different resolutions: The red solid line shows the results for the resolution used for all the simulations in this work $(n_{\theta}, n_{r})=(128,220)$ and the dashed, blue lines displays the quantities for the higher resolution of $(n_{\theta}, n_{r})=(256,460)$.}
    \label{fig:E_expl_evol_higher_res_comp}
\end{figure}

\subsection{Resolution study}\label{Appendix:resolution_study}

The resolution of the computational domain we use in this work has been chosen considering its effect on the duration of the simulations and the convergence of the results. In order to verify that $(n_{\theta}, n_{r})=(128,220)$ is a sufficient discretization of our domain, for M20\_10\_1\_0.3\_0.1 we also performed a simulation increasing the resolution to $(n_{\theta}, n_{r})=(256,460)$. Fig.~\ref{fig:M_disk_M_inj_higher_res_comp} shows the disk mass and the injected mass measured in the two simulations. The difference of these quantities in the two runs ranges from $\sim 0.01\%$ to $\sim 2.9\%$, confirming that using resolution of $(n_{\theta}, n_{r})=(128,220)$ we obtain converged results. The good agreement of the two simulations is also visible comparing the evolution of the explosion energy, as done in Fig.~\ref{fig:E_expl_evol_higher_res_comp}.

\begin{table*}
\centering
\caption{Model description and key results for models with an increased frequency (every 10~ms) of output. From left to right, the columns contain wind time scale, the ratio of the accretion and wind time scales, the squared ratio of the asymptotic velocity of injected matter to escape velocity of the disk, the internal to kinetic energy ratio of injected matter, cumulative injected energy, ejecta mass, explosion energy, average ejecta velocity, the mass of ejecta component that is originated from the computational domain and experienced temperature higher than 5\,GK, the mass of the $^{56}$Ni synthesized, the mass of ejecta component originated from the injected matter.
}
\begin{tabular}{lccccccccccc}
\hline\hline
model &$t_\mathrm{w}$ & $t_\mathrm{acc}/t_\mathrm{w}$ & $\xi^2$ & $f_\mathrm{therm}$ & $E_\mathrm{inj}$ & $M_\mathrm{ej}$ & $E_\mathrm{expl}$ & $v_\mathrm{ej}$ & $M_\mathrm{ej,>5GK}^\mathrm{stellar}$ & $M_\mathrm{ej,Ni}^\mathrm{stellar}$ & $M_\mathrm{ej}^\mathrm{inj}$ \\ %& $M_\mathrm{Ni}^\mathrm{inj}$ \\
&(s) & & & & ($10^{51}$\,erg) & ($M_\odot$) & ($10^{51}$\,erg) & ($10^3$\,km/s) & ($M_\odot$) & ($M_\odot$) & ($M_\odot$) \\ %& ($M_\odot$) \\ 
\hline
M20\_10\_3.16\_0.1\_0.10\_hf &    10 &     3.16 & 0.1 & 0.10 &      12 &   4.4 &   7.7 &    13 &   0.060 &   0.036  &       1.3 \\
M20\_10\_10\_0.1\_0.10\_hf &    10 &       10 & 0.1 & 0.10 &      15 &   5.3 &    11 &    14 &   0.069 &   0.039 &       2.0 \\
M20\_10\_inf\_0.1\_0.10\_hf &    10 & $\infty$ & 0.1 & 0.10 &      17 &   5.9 &    12 &    14 &   0.090 &   0.056 &       2.0  \\
M20\_3.16\_3.16\_0.3\_0.10\_hf &3.16 &     3.16 & 0.3 & 0.10 &      14 &   5.4 &   8.0 &    12 &  0.071 &  0.059 & 0.97 \\
M20\_3.16\_10\_0.3\_0.10\_hf &3.16 &       10 & 0.3 & 0.10 &      15 &   5.8 &   9.2 &    13 &   0.071 &   0.065 & 0.87 \\
M20\_3.16\_inf\_0.3\_0.10\_hf &3.16 & $\infty$ & 0.3 & 0.10 &      16 &   6.0 &    11 &    13 &   0.081 & 0.068 &       1.2 \\
M20\_10\_1\_0.3\_0.10\_hf &  10 &        1 & 0.3 & 0.10 &      19 &   4.6 &    12 &    16 &   0.050 &  0.045 & 0.70 \\
M20\_10\_3.16\_0.3\_0.10\_hf &  10 &     3.16 & 0.3 & 0.10 &      32 &   5.9 &    25 &    21 &   0.072 &  0.073 & 0.99 \\
M20\_10\_10\_0.3\_0.10\_hf &  10 &       10 & 0.3 & 0.10 &      38 &   6.6 &    32 &    22 &     0.18 &  0.15 &    1.6 \\
M20\_10\_inf\_0.3\_0.10\_hf &  10 & $\infty$ & 0.3 & 0.10 &      40 &   6.8 &    34 &    23 &  0.11 &  0.091 &      1.6 \\
M20\_3.16\_1\_0.1\_0.01\_hf & 3.16 &        1 & 0.1 & 0.01 &     1.4 &  0.58 & 0.053 &   3.0 & 0.0099 & 0.012 & 0.0081 \\
M20\_3.16\_3.16\_0.1\_0.01\_hf & 3.16 &     3.16 & 0.1 & 0.01 &     7.3 &   2.8 &   2.5 &   9.6 & 0.081 &   0.044 &    0.61 \\
M20\_3.16\_10\_0.1\_0.01\_hf & 3.16 &       10 & 0.1 & 0.01 &     8.6 &   3.9 &   4.5 &    11 &  0.086 &  0.050 &      1.2 \\
M20\_3.16\_inf\_0.1\_0.01\_hf & 3.16 & $\infty$ & 0.1 & 0.01 &     9.5 &   4.3 &   5.0 &    11 &    0.11 &  0.066 &   1.2 \\
M20\_10\_3.16\_0.1\_0.01\_hf &   10 &     3.16 & 0.1 & 0.01 &      12 &   5.3 &   8.2 &    12 &   0.082 &  0.049 &      1.9 \\
M20\_10\_10\_0.1\_0.01\_hf &  10 &       10 & 0.1 & 0.01 &      14 &   5.7 &   9.9 &    13 &  0.090 &   0.049 &     2.0 \\
M20\_10\_inf\_0.1\_0.01\_hf &   10 & $\infty$ & 0.1 & 0.01 &      17 &   5.9 &    12 &    14 &  0.092 &  0.058 &      2.3\\
\hline
\end{tabular}
\label{tab:models_hf}
\end{table*}

\subsection{Dependence of particle tracing on time interval of outputs}\label{Appendix:output_dependence_particle_tracing}
As mentioned in Section~\ref{sec:Nucleosynthesis}, the accuracy of the thermodynamical history evaluated in the post-process particle tracing is strongly affected by the frequency of the output. For most of the model we use a time interval of the output of 70 ms. We check the systematic performing the particle tracing with an output frequency increased to every 10 ms for selected model.
The results obtained for $M^\mathrm{stellar}_\mathrm{ej,>5GK}$ and  $M^\mathrm{stellar}_\mathrm{ej,Ni}$ with the increased frequency are displayed in parenthesis Table~\ref{tab:models_hf}. By comparing them with the results obtained for the same models but with lower output frequency listed in Table~\ref{tab:models}, they show that the uncertainty on the accuracy of the particle tracing is of $10$\% level.

%%%%%%%%%%%%%%%%%%%%%%%%%%%%%%%%%%%%%%%%%%%%%%%%%%

% Don't change these lines
%\bsp	% typesetting comment
\label{lastpage}
\end{document}